\documentclass[runningheads]{llncs}

\usepackage[T1]{fontenc}

\usepackage[ruled,vlined,linesnumbered]{algorithm2e}
\usepackage{hyperref}
\usepackage{amsmath,amssymb}
\usepackage{pgfplots}
\usepackage{graphicx}
\usepackage{caption}
\usepackage{subcaption}
\usepackage{mathtools}
\usepackage{multirow}
\usepackage{booktabs}
\usepackage{enumitem}

\allowdisplaybreaks

{}

\newtheorem{assumption}{Assumption}

\newcommand{\predbarrier}{\mu}
\newcommand{\statee}{\mathbf{s}}
\newcommand{\timeid}{k}
\newcommand{\states}{\mathcal{S}}
\newcommand{\controls}{\mathcal{C}}
\newcommand{\reals}{\mathbb{R}}

\newcommand{\integers}{\mathbb{Z}}

\newcommand{\pposint}{\integers^{> 0}}
\newcommand{\dynamics}{\mathbf{f}}
\newcommand{\action}{\mathbf{a}}

\newcommand{\control}{\mathbf{u}}

\newcommand{\mypara}[1]{\vspace{0.8em} \noindent {\bf #1}.}
\newcommand{\myipara}[1]{\vspace{0.8em} \noindent {\em #1}.}
\newcommand{\traj}{\sigma}
\newcommand{\init}{\mathcal{I}}

\def\bbetta{\boldsymbol{\beta}}
\def\balphha{\boldsymbol{\alpha}}

\newcommand{\horizon}{K}

\newcommand{\alw}{\mathbf{G}}
\newcommand{\ev}{\mathbf{F}}
\newcommand{\unt}{\mathbf{U}}
\newcommand{\intvl}{I}

\newcommand{\relu}{\mathbf{ReLU}}
\newcommand{\NN}{\eta}

\newcommand{\stltonn}{\mathsf{STL2NN}}
\newcommand{\trpzmffnn}{\mathcal{R}_\varphi}

\newcommand{\norm}[1]{\left\| #1 \right\|}

\newcommand{\horiz}{\mathsf{H}}
\newcommand{\ignore}[1]{}

\newcommand{\release}{\mathbf{R}}
\newcommand{\rob}{\rho}

\begin{document}

\title{A Neurosymbolic Approach to the Verification of Temporal Logic Properties of Learning enabled Control  Systems}
%
\titlerunning{Verification of Temporal Logic Properties}

\author{Navid Hashemi\inst{1}  \and Bardh Hoxha \inst{2} \and Tomoya Yamaguchi \inst{2}   \and Danil Prokhorov \inst{2} \and  Geogios Fainekos \inst{2} \and Jyotirmoy Deshmukh \inst{1} }
\authorrunning{N. Hashemi et al.}
\institute{University of Southern California, Los Angeles CA, USA \and
Toyota Research Institute North America, Ann Arbor, MI, USA\\
\email{jdeshmuk@usc.edu} }

\maketitle

\begin{abstract}
Signal Temporal Logic (STL) has become a popular tool for expressing formal requirements of Cyber-Physical Systems (CPS).  The problem of verifying STL properties of neural network-controlled CPS remains a largely unexplored problem. In this paper, we present a model for the verification of Neural Network (NN) controllers for general STL specifications using a custom neural architecture where we map an STL formula into a feed-forward neural network with ReLU activation. In the case where both our plant model and the controller are ReLU-activated neural networks, we reduce the STL verification problem to reachability in ReLU neural networks. We also propose a new approach for neural network controllers with general activation functions; this approach is a sound and complete verification approach based on computing the Lipschitz constant of the closed-loop control system. We demonstrate the practical efficacy of our techniques on a number of examples of learning-enabled control systems.

\end{abstract}
\keywords{ Signal Temporal Logic, Verification, Deep Neural Network, Lipstchitz constant, Reachability, Model, Controller, $\relu$  }

\section{Introduction}
Learning-enabled components (LECs) offer the promise of data-driven
control, and hence they are becoming popular in many Cyber-physical system, CPS, applications. Among LECs, controllers trained using deep
learning are becoming popular due to the advances in techniques like
deep reinforcement learning and deep imitation learning.  On one hand,
the use of such LECs has the potential of achieving human level
decision making in tasks like autonomous driving, aircraft collision
avoidance, and control for aerial vehicles. 
On the other hand, the use of deep neural network (DNN)-based
controllers raises serious concerns of safety. 

Reasoning about DNNs is
a challenge because DNNs are highly nonlinear
\cite{szegedy2013intriguing}, and
due to the nature of data-driven control, the behavior of a DNN
controller at a previously unseen state can be difficult to predict \cite{liu2021algorithms}.
To address this challenge, there has been significant research on {\em
verification} for DNNs. Broadly, there are two categories of
verification methods; the first category considers DNN controllers in
isolation and reasons about properties such as input-output robustness
\cite{ehlers2017formal,huang2017safety,katz2017reluplex}, range
analysis \cite{dutta2017output}, symbolic constraint propagation
through DNNs \cite{li2019analyzing}, and overapproximate reachable set computation
for DNNs \cite{tran2020nnv}. The second category of methods reasons
about DNN controllers in closed-loop with a dynamical model of the
environment/plant \cite{dutta2017output,huang2019reachnn,huang2021polar,ivanov2019verisig}. 

In this paper, we also address the closed-loop verification problem.
In this problem, we are typically provided with a set of inital states
and a set of unsafe states for the system, and the goal is to prove
that starting from an arbitrary initial state, no system behavior ever
reaches a state in the unsafe set. 
However, we extend this
problem in a significant manner. First, we assume that the desired
behavior of the closed-loop system is specified as a bounded horizon
Signal Temporal Logic (STL) \cite{maler2004monitoring} formula.
Second, in contrast to most existing closed-loop verification methods that
typically assume that an analytic representation of the system dynamics exists,
we allow the system dynamics themselves to be represented as a DNN.
Such a setting is quite common in techniques such as model-based deep
reinforcement learning \cite{chua2018deep,deisenroth2013gaussian}. This 
crucially allows us to reason about systems where the
analytic representation of the system dynamics may not
be available.

The central idea in our paper is a neurosymbolic verification
approach: we reformulate the robust satisfaction (referred to as robustness) of an
STL formula w.r.t. a given trajectory as a feed-forward neural network
with $\relu$ activation functions.
We call this transformation $\stltonn$. 
We show that
the output of $\stltonn$ is positive iff the STL formula is satisfied by
the trajectory. We note that the verification problem 
only requires establishing that the given closed-loop
dynamical system {\em satisfies} a given STL 
specification. However, by posing the verification problem
as that of checking robust satisfaction, it allows us to
conclude that the given DNN controller {\em robustly}
satisfies the given specification.

We then show that when the DNN-controller uses
$\relu$ activation functions, the problem of closed-loop STL
verification can be reduced to computing the reachable set for a
$\relu$-DNN. If the controller is not a $\relu$ neural network, we
propose a technique called Lip-Verify based on computing the Lipschitz
constant of the robustness of the given STL formula (as a function of
the initial state).

To summarize, the main contributions in this paper are:
\begin{enumerate}

\item
We formulate a neuro-symbolic approach for the closed-loop
verification of a DNN-controlled dynamical system against an STL-based
specification by converting the given bounded horizon 
specification into a feed-forward $\relu$-based DNN that we call
$\stltonn$.

\item
For data-driven plant models using $\relu$ activation and
$\relu$-activation based DNN-controllers, we show that the
verification of arbitrary bounded horizon STL properties can be
reduced to computing the reach set of the composition of the 
plant and controller DNNs with $\stltonn$.

\item

For arbitrary nonlinear plant models\footnote{In the experimental
results, we focus on linear and DNN plant models, but our method is 
applicable to other nonlinear plant models as well.} and DNN-controllers 
using arbitrary activation functions, we compute Lipschitz constant of the
function composition of the system dynamics with STL robustness, and
use this to provide a sound verification result using systematic
sampling.

\end{enumerate}

The rest of this paper is as follows. In Section \ref{sec:prelim}, we
present the background, primary concepts with STL semantics and
problem definition. In Section \ref{sec:STLNN}, we present the steps
to characterize $\stltonn$. In Section \ref{sec:Ver} we classify the
verification problem based on the involved activation functions and
propose a verification method for each class. We also introduce a
structure for formulation of verification problems and introduce our
verification toolbox. Finally, we present several case studies and
experimental results for our verification methods in Sections
\ref{subsec:reluresults} and \ref{subsec:lipverifyresults}. We  conclude with a discussion on related works
in Section \ref{sec:conclusion}.

\section{Preliminaries}
\label{sec:prelim}
In this section, we first provide the mathematical notation and
terminology to formulate the problem definition.  We use bold letters
to indicate vectors and vector-valued functions, and calligraphic
letters to denote sets. We assume that the reader is familiar with
feedforward neural networks, see \cite{goodfellow2016deep} for a
brief review.

\mypara{Neural Network Controlled Dynamical Systems (NNCS)}
Let $\statee$ and $\control$ respectively denote the state and input
control variables that take values from compact sets $\states
\subseteq \reals^n$ and $\controls \subseteq \reals^m$, respectively.
We use $\statee_\timeid$ (resp. $\control_\timeid$) to denote the
value of the state variable (resp. control input) at time $\timeid$.
We first define deep neural network controlled systems (NNCS) as a
recurrent difference equation\footnote{We note that in some modeling
scenarios, the dynamical equation describing the environment may be
provided as continuous-time ODEs. In this case, we assume that
we can obtain a difference equation (through numerical approximations
such as a zero-order hold of the continuous dynamics).  Our
verification results are then applicable to the resulting
discrete-time approximation. Reasoning about behavior between sampling
instants can be done using standard error analysis arguments that we do not consider
in this paper \cite{atkinson2011numerical}.}:
\begin{equation}
\label{eq:nncs}
    \statee_{\timeid+1} = \dynamics(\statee_\timeid,\control_\timeid), 
    \quad 
    \control_{\timeid} = \NN(\statee_\timeid).
\end{equation}
Here, $\dynamics$ is assumed to be any computable function, and $\NN$
is a (deep) neural network. We note that we can include time as a
state, which allows us to encode time-varying plant models as well
(where the dynamics corresponding to the time variable simply
increment it by $1$).


\mypara{Neural Plant Models} In the model-based development paradigm,
designers typically create environment or plant models using laws of
physics. However, with increasing complexity of real world
environments, the data driven control paradigm suggests the use of
machine learning models like Gaussian Process \cite{rasmussen2003gaussian} or neural
networks as function approximators. Such models typically take as
input the values of the state and control input variables at time
$\timeid$ and predict the value of the state at time $\timeid+1$. In
this paper, we focus on environment models that use deep 
neural networks\footnote{As we see later, the STL verification
technique that we formulate is compatible with using plant models that
use standard nonlinear functions, e.g. polynomials, trigonometric
functions, etc. However this requires integrating our method with
closed-loop verification tools such as Polar \cite{huang2021polar} , Sherlock
\cite{dutta2017output} , or NNV \cite{tran2020nnv} . We will consider this integration in the
future.}. 
On the other hand linear time-invariant (LTI) models can be
considered as a neural network with only linear activation functions.
Finally, we note that
our technique can also handle {\em time-varying} plant 
models such as linear time-varying models and DNN plant models that
explicitly include time as an input. 



\mypara{Closed-loop Model Trajectory, Task Objectives, and Safety
Constraints} Given a discrete-time NNCS as shown in
\eqref{eq:nncs}, we define $\init \subseteq \states$ as a set
of initial states of the system.  For a given initial state
$\statee_0$, and a given finite time horizon $\horizon \in \pposint$,
a system trajectory $\traj_{\statee_0}$ is a function from
$[0,\horizon]$ to $\states$, where $\traj_{\statee_0}(0) = \statee_0$,
and for all $\timeid \in [0,\horizon-1]$,
$\traj_{\statee_0}(\timeid+1) =
\dynamics(\statee_\timeid,\NN(\statee_\timeid))$.
We assume that task objectives or safety constraints of the system are
specified as bounded horizon Signal Temporal Logic (STL) formulas
\cite{maler2004monitoring}; the syntax\footnote{We do not include
the negation operator as it is possible to rewrite any STL formula in
negation normal form by pushing negations to the signal predicates
\cite{ho2014online}} of STL is as defined in
Eq.~\eqref{eq:stlfrag}.
\begin{eqnarray}
\label{eq:stlfrag}
\varphi ::=  \predbarrier(\statee) \bowtie 0 \mid 
             \varphi \wedge \varphi \mid 
             \varphi \vee \varphi \mid 
             \ev_\intvl \varphi \mid 
             \alw_\intvl \varphi \mid 
             \varphi_1 \unt_\intvl \varphi_2  \mid
             \varphi_1 \release_\intvl \varphi_2
\end{eqnarray}
\noindent Here, $\predbarrier$ is a function representing a {\em
linear combination} of $\states$ that maps to a number in $\reals$,
$\bowtie \in \{ <, \le, >, \ge \}$ and $\intvl$ is a compact interval
$[a,b] \subseteq [0,\horizon]$. The temporal scope or horizon of an
STL formula defines the number of time-steps required in a trajectory
to evaluate the formula. The horizon $\horiz(\varphi)$ of an STL
formula $\varphi$ can be defined as follows:
\[
\left\{
\begin{array}{ll}
0 & \text{if $\varphi \equiv \predbarrier(\statee) \bowtie 0$} \\
\max(\horiz(\varphi_1),\horiz(\varphi_2)) & 
    \text{if $\varphi \equiv \varphi_1 \circ \varphi_2$, 
    where $\circ \in \{\wedge,\vee\}$} \\ 
b + \horiz(\psi) & \text{if $\varphi = \mathbf{Q}_{[a,b]} \psi$, 
    where $\mathbf{Q} \in \{ \alw,\ev \}$} \\
b + \max(\horiz(\varphi_1),\horiz(\varphi_2)) &
    \text{if $\varphi = \varphi_1 \mathbf{Q}_{[a,b]} \varphi_2$, 
    where $\mathbf{Q} \in \{ \unt, \release \}$}
\end{array}\right.
\]

\mypara{Quantitative Semantics of STL} The Boolean semantics of STL
define what it means for a trajectory to satisfy an STL formula. A
detailed description of the Boolean semantics can be found in
\cite{maler2004monitoring}. The quantitative semantics of STL define
the signed distance of the trajectory from the set
of traces satisfying or violating the formula.  This signed distance
is called the {\em robustness} value. There are a number of ways to
define the quantitative semantics of STL
\cite{donze2010robust}, \cite{fainekos2006robustness}, \cite{rodionova2022combined}, \cite{akazaki2015time}; in
this paper, we focus on the semantics from \cite{donze2010robust} that
we reproduce below. The robustness value 
$\rob(\varphi,\traj_{\statee_0},\timeid)$ of an STL formula $\varphi$
over a trajectory $\traj_{\statee_0}$ at time $\timeid$ can be defined
recursively as follows. For brevity, we omit the trajectory from the
notation as it is obvious from the context.
\begin{equation}
\label{eq:stlrob}
\begin{array}{l|l}
\varphi & \rob(\varphi,\timeid) \\
\hline
\predbarrier(\statee) \ge 0 &
\predbarrier(\statee_\timeid) \\
\varphi_1 \wedge \varphi_2 &
\min(\rob(\varphi_1,\timeid),\rob(\varphi_2,\timeid)) \\
\varphi_1 \vee \varphi_2 &
\max(\rob(\varphi_1,\timeid),\rob(\varphi_2,\timeid)) \\
\alw_{[a,b]}\psi &
\min_{\timeid' \in [\timeid+a,\timeid+b]} \rob(\psi,\timeid) \\
\ev_{[a,b]}\psi &
\max_{\timeid' \in [\timeid+a,\timeid+b]} \rob(\psi,\timeid) \\
\varphi_1 \unt_{[a,b]} \varphi_2 &
\displaystyle\max_{\timeid' \in [\timeid+a,\timeid+b]}\left(
    \min\left(\begin{array}{l}
         \rob(\varphi_2,\timeid'), \\
         \min_{\timeid'' \in [\timeid,\timeid')} \rob(\varphi_1,\timeid'')
              \end{array}\right)
    \right) \\
\varphi_1 \release_{[a,b]} \varphi_2 &
\displaystyle\min_{\timeid' \in [\timeid+a,\timeid+b]}\left(
    \max\left(\begin{array}{l}
         \rob(\varphi_1,\timeid'), \\
         \max_{\timeid'' \in [\timeid,\timeid']} \rob(\varphi_2,\timeid'')
              \end{array}\right)
    \right) \\
\end{array}
\end{equation}
We note that if $\rob(\varphi,\timeid) > 0$ the STL formula
$\varphi$ is satisfied at time $\timeid$ (from
\cite{fainekos2006robustness}).

\mypara{Problem Definition}
The STL verification problem can be formally stated as follows:
Given an NNCS as shown in \eqref{eq:nncs}, a set of initial
conditions $\init$, and a bounded horizon STL formula $\varphi$ with
$\horiz(\varphi) = \horizon$, show that:
\begin{equation}
\label{eq:stlver}
\forall \statee_0 \in \init: \rob(\varphi, \traj_{\statee_0}, 0) > 0,
\end{equation}
\noindent where, the time horizon for $\traj_{\statee_0}$ is
$\horizon$.

%

\section{STL Robustness as a Neural Network}
\label{sec:STLNN}
In this section, we describe how the robustness of a bounded horizon
STL specification $\varphi$ with horizon = $\horizon$ over a
trajectory of length $\horizon$ can be encoded using a neural network
with $\relu$ activation functions. The first observation is that the
quantitative semantics of STL described in \eqref{eq:stlrob} can be
recursively unfolded to obtain a tree-like representation where the
leaf nodes of the tree are evaluations of the linear predicates at
various time instants of the trajectory and non-leaf nodes are $\min$
or $\max$ operations.  The second observation is that $\min$ and
$\max$ operations can be encoded using a $\relu$ function. We codify
these observations in the following lemmas.

\begin{lemma}\label{lem:reluminmax} Given $x,y \in \reals$, $\min(x,y) = W_2\cdot\relu (W_1\cdot[x\ \ y]^T)$,
where $W_1$ and $W_2$ are as given below. Similarly, 
$\max(x,y) = W_1\cdot\relu(W_3\cdot[x\ \ y]^T)$, where $W_3$ is as
given below.
\begin{equation}
W_1 = \begin{bmatrix}
       1 & 1  \\
      -1 & -1 \\
       1 & -1 \\
      -1 &  1 \\
      \end{bmatrix} %
W_2 = {\begin{bmatrix}
      0.5  \\
      -0.5 \\
      -0.5 \\
      -0.5 \\
      \end{bmatrix}}^T %
W_3 = {\begin{bmatrix}
       0.5 \\
      -0.5 \\
       0.5 \\
       0.5 
       \end{bmatrix}}^T
\end{equation}
\end{lemma}
\proof  We only provide the proof for the $\min(x,y)$, the proof for $\max$
follows symmetrically. Recall that for $\mathbf{v} \in \reals^m$,
$\relu(\mathbf{v}) = \max(\mathbf{v},\mathbf{0})$, i.e., a column
vector (say $\mathbf{r}$) of length $m$ where $\mathbf{r}(j) =
\max(\mathbf{v}(j),0)$. Consider the expression
$W_2\cdot\relu(W_1\cdot[x\ \ y]^T)$. The inner matrix multiplication
evaluates to:
\[\begin{bmatrix}x+y & -x-y & x-y &-x+y\end{bmatrix}^T.\]
Performing $\relu$ on this matrix will return one of four column
vectors (denoted $W'$): 
\[
\begin{array}{ll}
\begin{bmatrix}x+y & 0 & x-y & 0\end{bmatrix}^T & x+y \ge 0, x \ge y \\
\begin{bmatrix}x+y & 0 & 0 & y-x\end{bmatrix}^T & x+y \ge 0, y \ge x \\
\begin{bmatrix}0 & -x-y & x-y & 0\end{bmatrix}^T & x+y \le 0, x \ge y \\
\begin{bmatrix}0 & -x-y & 0 & y-x\end{bmatrix}^T & x+y \le 0, y \ge x \\
\end{array}
\]
Now, consider the outer multiplication, $W_2.W'$. This multiplication
will result in one of four values (depending on which case above is
true):  $0.5(x+y) + -0.5(x-y) = y$ (when $x \ge y$), 
$0.5(x+y) + -0.5(y-x) = x$ (when $y \ge x$),
$-0.5(-x-y) + -0.5(x-y) = y$ (when $x \ge y$),
$0.5(-x-y) + -0.5(y-x) = x$ (when $y \ge x$).  Note that irrespective of
the sign of $x+y$, the result of the multiplication always yields the
number that is $\min(x,y)$. \hspace*{\fill}~$\square$\\

\mypara{Mapping STL robustness to the $\stltonn$ neural network} We now
describe how to transform the robustness of a given STL formula and a
trajectory into a multi-layer network representation. Though we call
this structure a neural network, it is bit of a misnomer as there is
no learning involved. The name $\stltonn$ is thus reflective of the
fact that the structure of the graphical representation that we obtain
resembles a multi-layer neural network.

The input layer of $\stltonn$ is the set of all time points in the
trajectory (thus the input layer is of width $\horizon+1$). The second
layer is the application of the $m$ possible unique predicates
$\{\predbarrier_1,\ldots,\predbarrier_m\}$ in $\varphi$ to the
$(\horizon+1)$ possible time points. Thus, the ouptut of this layer is
of maximum dimension $m\times(\horizon+1)$. Let this layer be called the
predicate layer, and we denote each node in this layer by two
integers: $(i, \timeid)$, indicating the value of
$\predbarrier_i(\statee_\timeid)$.

For a trajectory $\traj_{\statee_0}$ of length $\horizon$, there are
at most $\horizon+1$ time points at which these $m$ predicates can be
evaluated. Thus, there are at most $m\times (\horizon+1)$ number of unique
evaluations of the $m$ predicates at $(\horizon + 1)$ time instants.


\SetKwProg{Fn}{Function}{}{end}
\newcommand{\mysty}[1]{$\mathtt{#1}$}
\SetKwSwitch{Switch}{Case}{Other}{switch}{}{case}{otherwise}{endcase}{endsw}
\SetFuncSty{mysty}
\SetKw{Return}{return}
\DontPrintSemicolon
\RestyleAlgo{ruled}
\begin{algorithm}[t]
\SetKwFunction{minmaxnode}{MinMaxNode}
\Fn{\minmaxnode{$n_1, \ldots, n_\ell$, $\mathsf{type}$, $\timeid$}}{
{\sf $\bullet$ Construct balanced binary tree with leafnodes $n_i,
     i=1,\ldots,\ell$.}\;
{\sf $\bullet$ Apply Lemma~\ref{lem:reluminmax} to obtain a $\relu$ network of depth 
     $O(log \ell)$ for $\min$ or $\max$ as defined by
     input type}
}
\SetKwFunction{networknode}{Node}
 \Fn{\networknode{$\varphi$,$\timeid$}}{
     \lCase{$\varphi = \predbarrier_i(\statee_\timeid) \ge 0$}{
        \Return $(i,\timeid)$  \nllabel{algoline:leafnode}
     }
     \uCase{$\varphi=\varphi_1 \wedge \varphi_2$}{
        \Return \minmaxnode\big(\networknode{$\varphi_1$,$\timeid$},\networknode{$\varphi_2$,$\timeid$}, $\mathsf{min}$, $\timeid$\big)
        \nllabel{algoline:conjunction}
     }
     \uCase{$\varphi=\varphi_1 \vee \varphi_2$}{
        \Return \minmaxnode\big(\networknode{$\varphi_1$,$\timeid$},\networknode{$\varphi_2$,$\timeid$}, $\mathsf{max}$, $\timeid$\big)
     }
     \uCase{$\varphi=\alw_{[a,b]}\varphi$}{
        \Return \minmaxnode 
        (\networknode{$\varphi$,$\timeid+a$}, $\ldots$ , \\
        \networknode{$\varphi$, $\timeid+b$}, $\mathsf{min}$,
        $\timeid$)
        \nllabel{algoline:alw}
     }
     \uCase{$\varphi=\ev_{[a,b]}\varphi$}{
        \Return \minmaxnode 
        (\networknode{$\varphi$,$\timeid+a$}, $\ldots$ , \\
        \networknode{$\varphi$, $\timeid+b$}, $\mathsf{max}$,
        $\timeid$)
        \nllabel{algoline:ev}
     }
     \uCase{$\varphi=\varphi_1\unt_{[a,b]}\varphi_2\ \mathsf{or}\ \varphi_1\release_{[a,b]}\varphi_2$}{
         {\sf similar to previous cases, following the robustness
         computation as defined in \eqref{eq:stlrob}}
     }
 }

\caption{Recursive formulation of a ReLU directed acyclic graph DAG for an STL formula}
\label{algo:networknode}
\end{algorithm}



Given the predicates $(i,\timeid)$ the Algorithm
\ref{algo:networknode} constructs the next segment of $\stltonn$.
Line~\ref{algoline:leafnode} returns the node corresponding to
$\predbarrier_i(\statee_\timeid)$, i.e. the node labeled $(i,\timeid)$
in the second layer of the network. Then the network structure follows
the structure of the STL formula. For example, in
Line~\ref{algoline:conjunction}, we obtain the nodes corresponding to
$\varphi_1$ and $\varphi_2$ at time $\timeid$, and these nodes are
then input to the $\relu$ unit that outputs the $\min$ of these two
nodes (as defined in Lemma~\eqref{lem:reluminmax}). The interesting
case is for temporal operators
(Lines~\ref{algoline:ev},\ref{algoline:alw}). A temporal operator
represents the $\min$ or $\max$ or combination thereof of subformulas
over different time instants. Suppose the scope of the temporal
operator requires performing a $\min$ over $\ell$ different time
instants, then in the function $\mathsf{MinMaxNode}$, we arrange these
$\ell$ inputs in a balanced binary tree of depth at most $O(\log
\ell)$ and repeatedely use the $\relu$ unit defined in
Lemma~\ref{lem:reluminmax} (see Appendix \ref{apdx:generalminmax}).
Executing Algorithm~\ref{algo:networknode}, will lead to a directed
acyclic graph, DAG network with depth at most $O(\log \horizon
|\varphi| )$ (as there are at most $|\varphi|$ operators in $\varphi$)
and each operator can require a network of depth at most $O(\log
\horizon)$. 

\mypara{DAG to feedforward NN} Algorithm~\ref{algo:networknode}
creates a DAG-like structure where nodes can be arranged in layers
(corresponding to the distance from the leaf nodes). However, this is
strictly not the structure of a feed-forward neural network as some
layers have connections that are skipped. To make the structure
strictly adhere to layer-by-layer computation, whenever an
$(i,\timeid)$ node is required in a deeper layer, we can add
neurons (corresponding to an identity function) that copy the value of
the $(i,\timeid)$ node to the next layer.  Observe that the addition
of these additional neurons does not increase the depth of the
network.  Thus, each layer in our $\stltonn$ has a mixture of
$\relu$-activation neurons and neurons with linear (identity)
activations.  We note that the position of these neurons corresponing to the linear
and $\relu$ activations can be separated through a process of
modifying the weight matrices for each layer. This separation of the
linear and $\relu$ layers is crucial in downstream verification
algorithms.  We call this neural network with redundant linear
activations and reordered neurons as $\stltonn$.  We codify the
argument for the depth of $\stltonn$ in Lemma~\ref{lemm:logdepth}. The
proof follows from our construction of $\stltonn$ in
Algorithm~\ref{algo:networknode}.

\begin{lemma}\label{lemm:logdepth}
Given a STL formula $\varphi$, the depth of $\stltonn$ increases logarithmically
with the length of the trajectory, $\traj_{\statee_0}$ and linearly in
the size of the formula.
\end{lemma}

\begin{theorem}\label{thm:main}
Given the STL formula, $\varphi$, the controller,
$\control_\timeid=\NN(\statee_\timeid)$ and the resultant trajectory
$\traj_{\statee_0}$,
\[
\rob(\varphi, \traj_{\statee_0}, 0)\geq 0 \iff \stltonn(\traj_{\statee_0}) \geq 0
\]
\end{theorem}

Lemma \ref{lemm:logdepth} shows, given a complex STL specification,
although the width of $\stltonn$ can be high, its depth is logarithmic
in the size of the trajectory,

\section{STL Verification using Reachability}
\label{sec:Ver}
In this section, we show how we can use the proposed $\stltonn$ for
verifying that a given STL formula $\varphi$ holds for {\em all}
initial states in a given set. Based on the structure of the plant
model and the kind of activation functions used by the DNN controller,
we will look at two different methods. We propose the overall
verification approach and a reachability analysis based sound and
complete method in this section. In the next section, we provide a
sampling-based sound and complete method.

\subsection{Trapezium feed-forward Neural Network (TNN)}
\label{sec:tnn}


\begin{figure}
    \includegraphics[width=\textwidth]{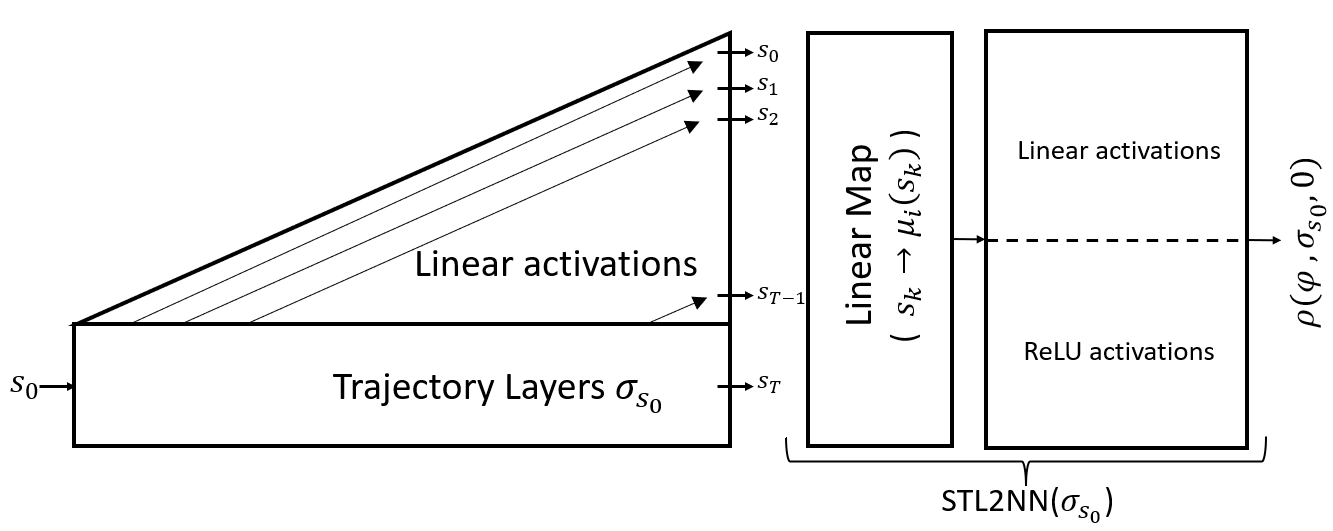}
    \caption{The structure of TNN (that encodes the computation of the
    trajectory $\traj_{\statee_0}$ starting from initial state
    $\statee_0$) composed with $\stltonn$ (that encodes the
    computation of the robustness of the STL formula $\varphi$ w.r.t.
    $\traj_{\statee_0}$).\label{fig:Trapezium}}
\end{figure}
Recall the dynamical system from \eqref{eq:nncs}, we can rewrite it
simply as $\statee_{\timeid+1} = \dynamics(\statee_\timeid,
\NN(\statee_\timeid))$. From this equation, we construct a neural
network that we call the {\em trapezium feed forward neural network}.
The name is derived from the shape in which we arrange the neurons.
The input to TNN is the initial state $\statee_0$. TNN has $\horizon$
blocks, where for $\timeid \ge 1$, the output of the
${\timeid-1}^{th}$ block is $\begin{bmatrix} \statee_0 & \cdots &
\statee_{\timeid-1} \end{bmatrix}$.  The $\timeid^{th}$ block
essentially takes the $\timeid$ outputs of the previous block and
``copies'' them to the block output using neuron layers that implement
identity maps. The $(\timeid+1)^{th}$ output of the block is the
computation of $\statee_{\timeid +1}$ using the difference equation
stated above. Thus, TNN has a shape where each subsequent block has
an equal number of additional number of neurons (equal to the
dimension of the state variable). The output of the $\horizon^{th}$
block can be then passed off to the input of $\stltonn$. Recall that
the output of $\stltonn$ is a single real number representing the
robustness value of $\varphi$ w.r.t. the trajectory
$\traj_{\statee_0}$.
We pictorially represent this 
in Fig.~\ref{fig:Trapezium}.
We remark that this structure is important and has a non-trivial
bearing on the verification methods that we develop in this paper as
we observe later. TNN thus encodes a function $\trpzmffnn:\reals^n \to
\reals$, where, 
\[
\trpzmffnn(\statee_0) = \rob(\varphi,\traj_{\statee_0}, 0).
\]

Given a TNN, we can use it to solve the problem outlined in
\eqref{eq:stlver}. In rest of this section, we show how we can use a
generic neural network reachability analyzer to perform STL
verification.

\subsection{STL verification using reachability analysis}

The following assumption encodes the fact that neural network
reachability analyzers are sound.
\begin{assumption}
\label{lem:nnreach}
Consider a neural network $N$ where the space of permitted inputs is
$X$. Then a neural network reachability analyzer produces as output a
set $Y$ s.t. $\forall x \in X: N(x) \in Y$. 
\end{assumption}

The following theorem establishes how we can reduce the problem of STL
verification to the problem of NN reachability.
\begin{theorem}
Given an NNCS as described in \eqref{eq:nncs}, a
set of initial conditions $\init \subseteq \states$, and a bounded
horizon STL
formula $\varphi$, we can reduce the problem of checking
\eqref{eq:stlver} to a NN reachability analysis problem.
\end{theorem}

\proof From Section~\ref{sec:tnn}, we know that given an NNCS, an initial
state $\statee_0$, and a bounded horizon STL formula $\varphi$, the
TNN function $\trpzmffnn$ encodes $\rob(\varphi,\traj_{\statee_0},0)$.
From Assumption~\ref{lem:nnreach}, if we have an NN reachability analyzer,
given the set $\init$ we can obtain a set (say $Y$) s.t. $\forall
\statee_0 \in \init: \trpzmffnn(\statee_0) \in Y$. We can then compute
$\inf Y$ and check if it is positive. If yes, the STL formula is
satisfied by the set of all initial conditions for the given NNCS.
\hspace*{\fill}~$\square$\\

While our method is broadly applicable with any NN reachability
analysis tool that can compute sound over-approximations of the set of
outputs for a given input set \cite{zhang2018efficient},\cite{dutta2017output}, \cite{huang2021polar},\cite{huang2019reachnn}, in this paper, we focus on a specific
type of NN reachability analysis tool that uses the notion of star
sets for performing reachability analysis \cite{tran2019star}. The
approach in \cite{tran2019star} performs exact reachability analysis
for DNNs with $\relu$ activation. 

Thus, for the star sets-based technique to be applicable, we require
that our {\em given plant model either uses $\relu$ activations or is
a linear model}, and {\em our controller uses $\relu$ activations}.
We can then apply star sets-based reachability by propagating the set
$\init$ through the TNN to compute the range of robustness values
through exact star based reachability analysis.  This verification is
sound and complete since the output range for $\trpzmffnn$ can be
accurately computed.\footnote{The TNN that we compute is a combination
of linear ($\mathtt{purelin}$) and $\relu$ activation functions. This implies
TNN is not a pure $\relu$ neural network, but the exact star set
reachability algorithm in \cite{tran2019star} can be updated to
include $\mathtt{purelin}$ activations and the exact reachability
analysis can be still performed on the TNN structure.}

Exact star based reachability can be time inefficient due to
exponential accumulation of star sets through reachability analysis
process. In this case, we can apply the approximate star based
technique \cite{tran2019star} to TNNs to perform verification.
Although this verification procedure is sound, it lacks completeness
as it may not be possible to algorithmically eliminate the
conservatism of the approximate reachability analysis.

\begin{remark}
We can also verify plant and controller models with arbitrary
activation functions using the TNN-based method, for example by using
NN reachability analysis tools for arbitrary activation functions,
such as the CROWN library \cite{zhang2018efficient}. However, we provide a sound and complete
verification procedure for NNs with arbitrary activation functions in
the next section.
\end{remark}

\subsection{Experimental Evaluation: STL verification with NN reachability}\label{subsec:reluresults}

In this section, we experimentally evaluate the efficacy of our
verification method and the TNN (and $\stltonn$) networks that we have
formulated. In the case studies considered in this section, we assume
access to the physics-based difference equations, which are used to
generate data to train the $\relu$-NN plant models. During training we
use the difference between the next state predicted by the DNN-based
plant model and the actual next state as the loss function. 
\begin{table*} [t]
\centering
\resizebox{\textwidth}{!}{
\begin{tabular}{ccccccccccccccc} 
\toprule
Reach &Property& $\init$ & Property & Model NN & Controller NN & Depth $\traj_{\statee_0}$ / $\stltonn$ & Robustness & Verified?& Run-time\\ 
Tech. & & & Horizon & structure & structure & & Range & & & \\
\midrule
E& $\varphi_1$ & $\init_1$ & 100 & [3,10,10,10,2]&[2,50,1,2,1,2,1,1]& 900\ /\ 15 \ layers & $\begin{bmatrix} 0.0150&0.0161 \end{bmatrix}$ & Yes & 1167 $\mathrm{sec}$\\ 
A & $\varphi_1$ & $\init_1$ & 100 & [3,10,10,10,2]&[2,50,1,2,1,2,1,1]& 900\ /\ 15 \ layers & $\begin{bmatrix} -0.0319&0.0256 \end{bmatrix}$ & No &35 $\mathrm{sec}$ \\ 
E& $\varphi_2$& $\init_2$ & 50 & [4,10,10,3]& [3,100,1,2,1,2,1,1]& 400\ /\  14 \ layers & $\begin{bmatrix} 0.0057630&0.005813 \end{bmatrix}$ & Yes & 1903 $\mathrm{sec}$\\ 
A &$\varphi_2$& $\init_2$ & 50 & [4,10,10,3]& [3,100,1,2,1,2,1,1]& 400\ /\ 14 \ layers & $\begin{bmatrix} -0.0308&0.0136 \end{bmatrix}$ & No &43 $\mathrm{sec}$ \\ 
E& $\varphi_3$& $\init_3$ & 53 & [7,10,10,6]& [5,20,20,20,1]& 265\ /\  9 \ layers & $\begin{bmatrix} 15.9077&38.4651 \end{bmatrix}$ & Yes & 259.7 $\mathrm{sec}$\\ 
A &$\varphi_3$& $\init_3$ & 53 & [7,10,10,6]& [5,20,20,20,1]& 265\ /\ 9 \ layers & $\begin{bmatrix} 11.6941&41.6572 \end{bmatrix}$ & Yes &23.82 $\mathrm{sec}$ \\
E& $\varphi_3$& $\init_3$ & 53 & [7,6] (LTI)& [5,20,20,20,1]& 159\ /\  9 \ layers & $\begin{bmatrix} 17.0904&38.9601 \end{bmatrix}$ & Yes & 139.4 $\mathrm{sec}$\\ 
A &$\varphi_3$& $\init_3$ & 53 & [7,6] (LTI)& [5,20,20,20,1]& 159\ /\ 9 \ layers & $\begin{bmatrix} 17.0904&38.9744 \end{bmatrix}$ & Yes &5.5 $\mathrm{sec}$ \\ 
E& $\varphi_4$& $\init_4$ & 32 & [4,8,2]& [2,8,2]& 64\ /\  10 \ layers & $\begin{bmatrix} 0.1033&0.2000 \end{bmatrix}$ & Yes & 77.78 $\mathrm{sec}$\\ 
E& $\varphi_5$& $\init_4$ & 35 & [4,8,2]& [2,8,2]& 70\ /\  16 \ layers & $\begin{bmatrix} 0.1033&0.1735 \end{bmatrix}$ & Yes & 1955 $\mathrm{sec}$\\ 
E& $\varphi_6$& $\init_4$ & 35 & [4,8,2]& [2,8,2]& 70\ /\  18 \ layers & $\begin{bmatrix} 0.1032&0.1462 \end{bmatrix}$ & Yes & 2368.8 $\mathrm{sec}$\\ 
E& $\varphi_7$& $\init_4$ & 36 & [4,8,2]& [2,8,2]& 72\ /\  18 \ layers & $\begin{bmatrix} -0.3271&0.1040 \end{bmatrix}$ & Rejected & 1023 $\mathrm{sec}$ \\
\bottomrule
\end{tabular} 
}
\caption{Shows the result of verification utilizing the reachability
analysis on TNN. In each case study we consider, both the plant model
and the controller are
$\relu$-FFNNs. We use the abbreviations \textbf{A} for Approximate star-set-based reachability, and \textbf{E} for the Exact star-set-based technique. No parallel computing is used and no set partitioning is applied.} \label{tbl:firstcategory}
\end{table*}

\begin{figure*}
  \begin{minipage}[c]{0.49\textwidth}
    \includegraphics[width=\linewidth]{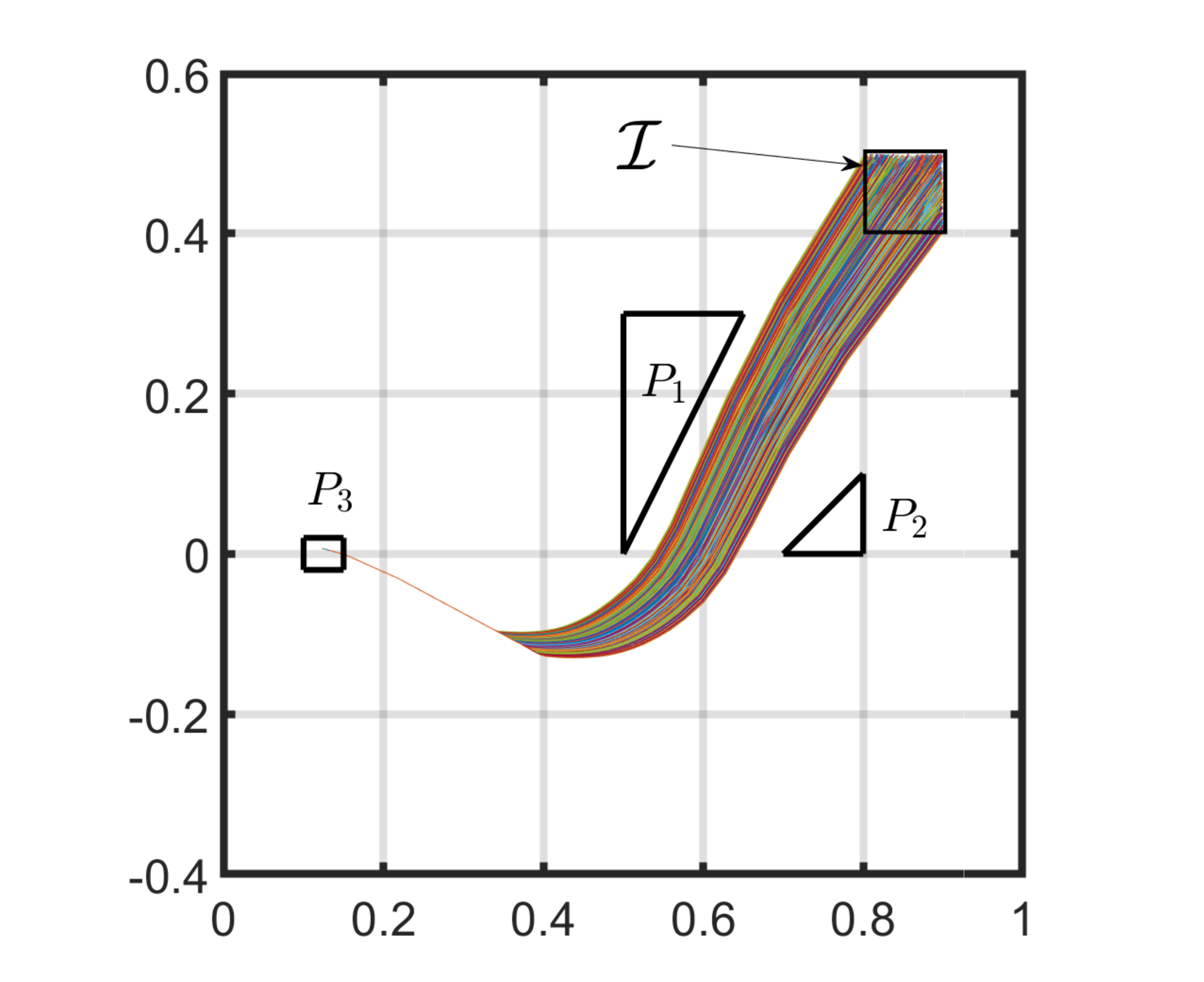}
    \captionof{figure}{\small{
     Trajectories for the model NFC-2d. The NN-controller is 
     required to drive trajectories to visit region $P_3$ within 
     time $\timeid$, where $\timeid \in [75,100]$. The controller
     should also avoid unsafe sets $P_1,\ P_2$ at all times.
    }} \label{fig:ex4}
  \end{minipage}\hfill
  \begin{minipage}[c]{0.49\textwidth}
    \includegraphics[width=\linewidth]{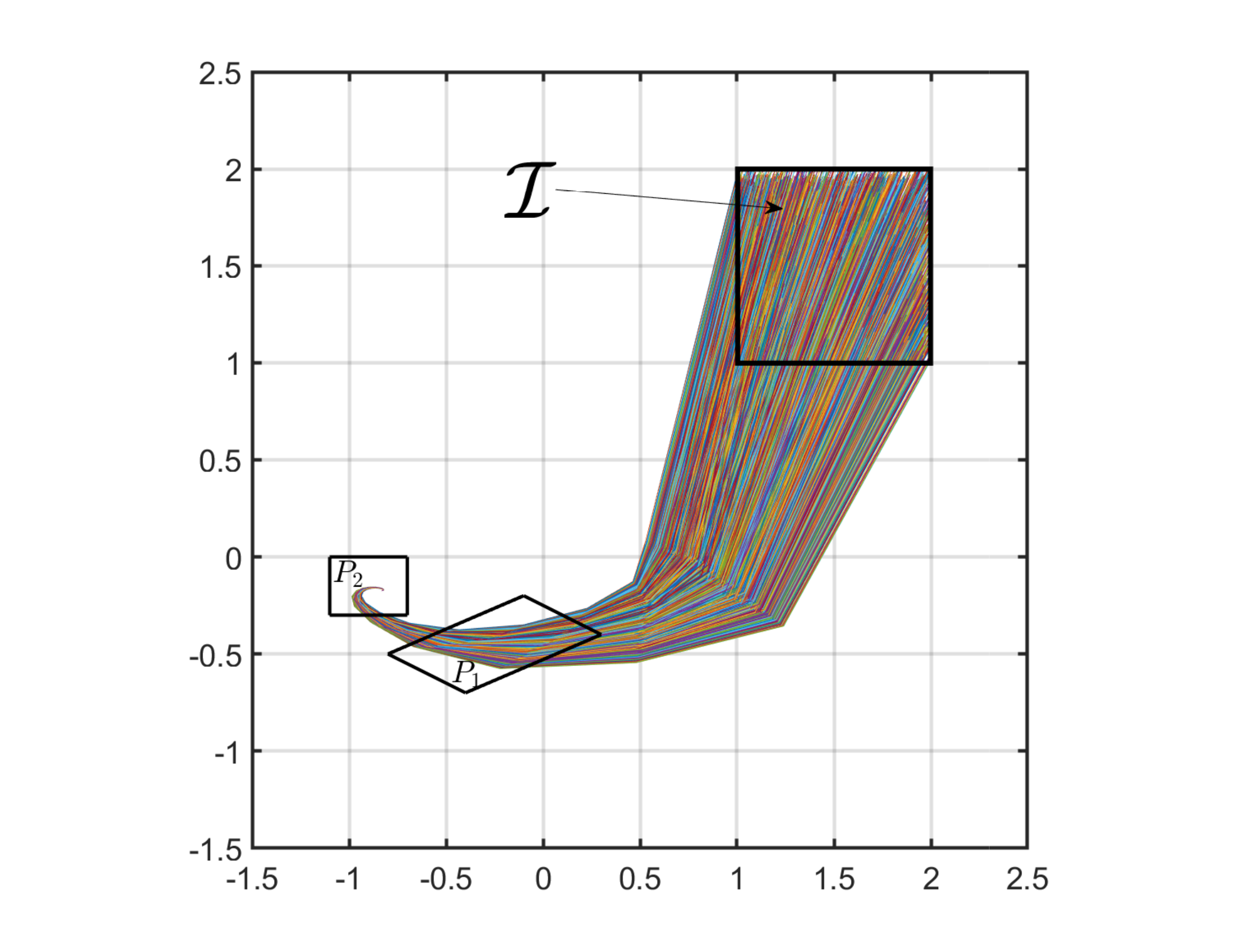}
\captionof{figure}{\small{Trajectories for the 
models using tangent-hyperbolic activation functions from Section~\ref{subsec:lipverifyresults}. The controller is required to drive the model such that it
 visits the region $P_1$ after $3$ time-steps but no
later than $6$ time-steps. Once it reaches region
$P_1$ it is required to visit $P_2$ after $9$ time
steps but no later than $13$ time steps.}} \label{fig:lip-example}
  \end{minipage}
\end{figure*}

\myipara{2D Nonlinear Feedback Control Model (NFC-2d)}: The symbolic
representation of the dynamics that was used to train the
$\relu$-plant model is shown in Eq.~\eqref{eq:nfc2d}, as the original
model is continuous-time, we used a sample time of $0.1$ seconds to
discretize the model before generating data. The number of samples,
i.e. the number of tuples of the form
($\statee_\timeid,\action_\timeid,\statee_{\timeid+1}$) that were used
in training was $10^6$.  The initial set of states $\init_1$ is as
shown in \eqref{eq:nfc2d}.  Figure \ref{fig:ex4} shows sample
trajectories of this model.%
\begin{equation}
\label{eq:nfc2d}
\resizebox{\hsize}{!}{$
\begin{bmatrix} \dot{x}_1 \\ \dot{x}_2 \end{bmatrix} = \begin{bmatrix}-x_1\left(0.1+(x_1+x_2)^2\right) \\ (u+x_1)\left(0.1+(x_1+x_2)^2\right) \end{bmatrix},\  \init_1=\left\{  \statee_0\ \mid \  \begin{bmatrix} 0.8\\0.4\end{bmatrix} \leq \statee_0 \leq \begin{bmatrix} 0.9\\0.5\end{bmatrix} \right\}
$}
\end{equation}
\noindent For this system, we are interested in verifying STL formula
$\varphi_1$ specified in Eq.~\eqref{eq:phi1}. This STL formula encodes
a classic ``reach-while-avoid'' specification of reaching region $P_3$
in a specific time-interval
\begin{equation}
\label{eq:phi1}
\varphi_1 = \ev_{[75,100]} \left(s\in P_3\right) \wedge 
            \alw_{[1,100]} \left(s \notin P_2 \right) \wedge 
            \alw_{[1,100]} \left( s \notin P_1 \right)
\end{equation}

\begin{figure}
    \centering
    \includegraphics[width=\linewidth]{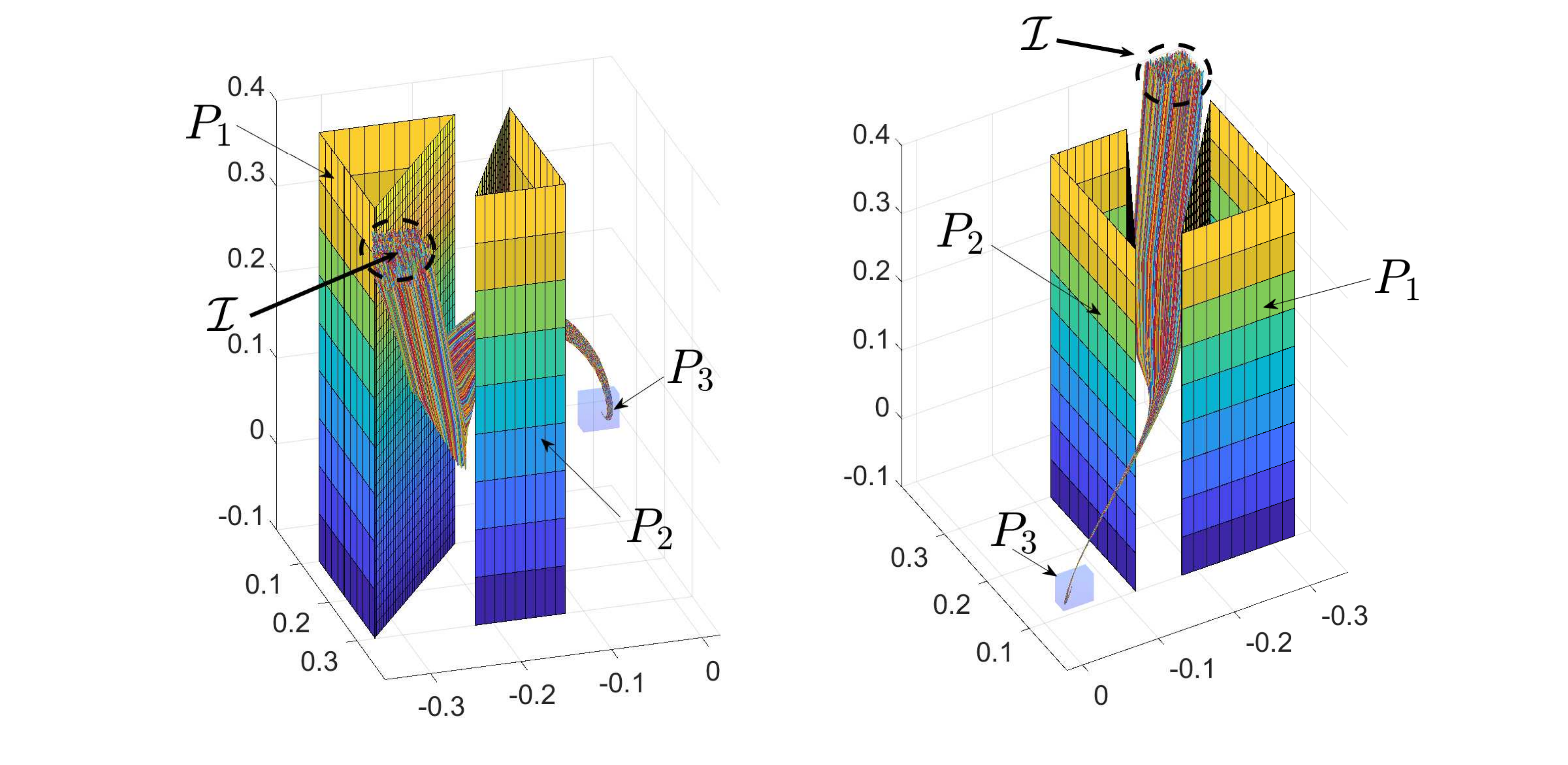}
    \caption{\small{
      Trajectories for the model NFC-3d. The NN-controller is 
      required to drive the model to the region $P_3$ within
      time $\timeid$, where $\timeid \in [35, 50]$, while 
      avoiding the unsafe sets $P_1,\ P_2$ at all times.
      }} \label{fig:ex7}
\end{figure}

\myipara{3D Nonlinear Feedback Control Model}
Figure.~\ref{fig:ex7} shows the trajectories of the nonlinear dynamical
model shown in Eq.~\eqref{eq:nfc3d}. The neural plant model is trained
on $1.35\times 10^6$ transitions after discretizing the model with a
sample time of $0.1$ seconds.
\begin{equation}
\label{eq:nfc3d}
\begin{bmatrix}\dot{x}_1\\ \dot{x}_2\\ \dot{x}_3\end{bmatrix}=\begin{bmatrix} x_1^3+x_2\\ x_2^3+x_3\\ u\end{bmatrix}, \  \init_2=\left\{  \statee_0\ \mid \  \begin{bmatrix} 0.35\\-0.35\\0.35\end{bmatrix} \leq \statee_0 \leq \begin{bmatrix} 0.4\\-0.3\\0.4\end{bmatrix} \right\}
\end{equation}
We want to verify if the controller satisfies the formula $\varphi_2$:
\begin{equation}
\label{eq:phi2}
\varphi_2 = \ev_{[35,50]} \left[ s \in P_3\right]  \bigwedge \alw_{[0,50]} \left[ s \not\in P_2\right] \bigwedge  \alw_{[0,50]} \left[ s \not\in P_1\right] 
\end{equation}

\myipara{Adaptive Cruise Control}
The third model we consider is a $\relu$-NN plant model fit to a 
discretization
of the $6$-dimensional adaptive cruise control model described in
Eq.~\eqref{eq:acc} (sample time was $0.1s$). We used $1.5\times 10^6$
samples to train the plant model.
In \eqref{eq:acc}, the constant $\mu$ denotes a
coefficient of friction set to $10^{-4}$. 
%
\begin{equation}
\label{eq:acc}
\begin{bmatrix}\dot{x}_1\\ \dot{x}_2\\ \dot{x}_3\\ \dot{x}_4\\
\dot{x}_5\\ \dot{x}_6\end{bmatrix}=\begin{bmatrix} x_2\\ x_3\\
-2x_3-4-\mu x_2^2\\ x_5\\ x_6\\ -2x_6+2u-\mu x_4^2\end{bmatrix}, \ \init_3=\left\{  \statee_0\ \middle| \  \begin{bmatrix} 90\\32\\0\\10\\30\\0\end{bmatrix} \leq \statee_0 \leq \begin{bmatrix} 110\\32.2\\0\\11\\30.2\\0\end{bmatrix} \right\}
\end{equation}
The NN-controller receives the observation, $O=[V_{set},\ t_{gap},\
x_5(\timeid),x_1(\timeid)-x_4(\timeid),\ x_2(\timeid)-x_5(\timeid)]$
and returns the optimal control to satisfy the proposed STL
specification~\eqref{eq:STL-ACC2} within $50$ time steps. Here
$V_{set} = 30$, and $t_{gap} = 1.4$ are fixed.
\begin{equation}\label{eq:STL-ACC2}
\resizebox{\hsize}{!}{$
\varphi_3=\alw_{[0,50]} \left(  \left[x_1(\timeid)-x_4(\timeid)<d_{safe} \right] \implies \ev_{[0,3]}\left[x_1(\timeid)-x_4(\timeid)>d_{safe}^* \right]\right)
$}
\end{equation}
where , $d_{safe}^*=12+1.4 x_5(\timeid)$ and $d_{safe}=10+1.4 x_5(\timeid)$. 

If the friction coefficient $\mu = 0$, then the model becomes an LTI
system, and we can perform STL verification of the NNCS (where the
plant has LTI dynamics).

\myipara{Model to test Scalability with respect to Complexity of STL Formula}

\begin{figure}
\includegraphics[width=\linewidth]{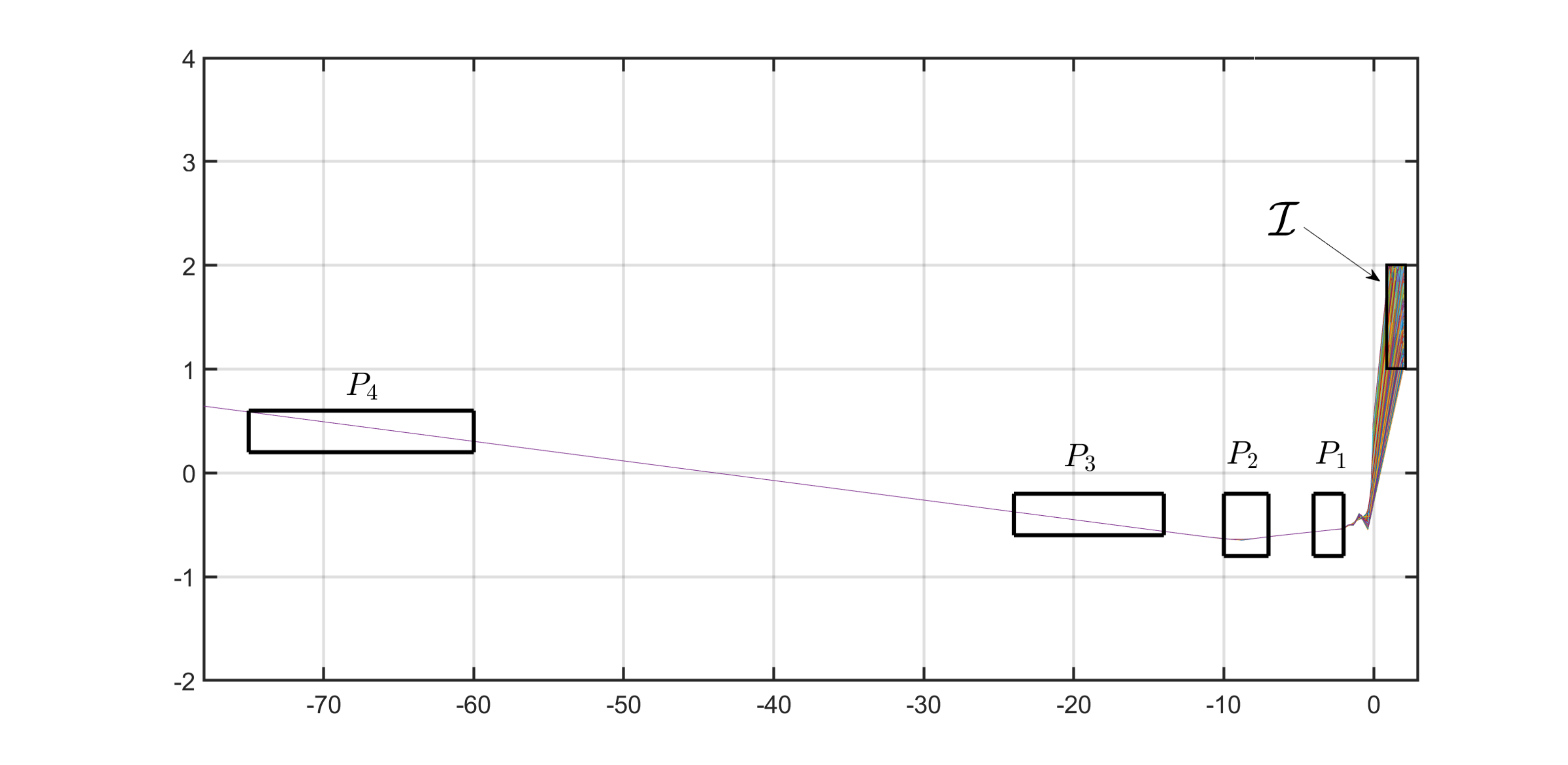}
\vspace{-5mm}
\caption{\small{Trajectories for the simple $\relu$-    
 plant and controller models used to show      scalability of STL verification in the  size of the STL formula. In this example we propose $4$ different sets and the moving object should visit them consecutively. We include them one by one in the STL property to make the formula more complex. The robustness upper bound is noticeably greater than lower bound. This implies although the trajectories look like they are very close in the figure, they have different  characteristics w.r.t. the STL formula.}} \label{fig:STLcomplex}
\end{figure}

To evaluate the scalability of our method with the complexity of the
STL specification, we constructed a simple $2D$ plant model and a
controller that takes $2$ inputs and produces $2$ outputs (both are
$\relu$-NNs). The actual model does not have any physical meaning (we
show the model trajectories in Appendix~\ref{sec:appendices} for
reference).  The initial set of states $\init_4$ is
$\left\{  \statee_0\ | \  \begin{bmatrix} 1&1\end{bmatrix}^\top \leq
\statee_0 \leq \begin{bmatrix} 2&2\end{bmatrix}^\top \right\}$.
We gradually increase the complexity of the STL spec to analyze the
runtime for verification using the exact-star reachability technique.
The STL formulas we use as verification targets are shown in
\eqref{eq:phi4}-\eqref{eq:phi6}. We want to show that the formula in 
\eqref{eq:phi7} is not satisfied by all initial states. The
difference in formula $\varphi_6$ and $\varphi_7$ is in the time
interval colored in red in $\varphi_7$.
{\footnotesize
\begin{eqnarray}
\varphi_4 & = & 
\ev_{[5,8]} \left( s \in P_1 \wedge \ev_{[20,24]}s \in P_4 \right) 
\label{eq:phi4} \\
\varphi_5 & = & 
\ev_{[5,8]} \left(  s \in P_1 \wedge  \ev_{[6,11]}\left( s \in P_2  \wedge \ev_{[12,16]}  s \in P_4   \right) \right) 
\label{eq:phi5}  \\
\varphi_6 & = & \ev_{[5,8]} \left( s \in P_1 \wedge  
                     \ev_{[6,11]}\left(s \in P_2 \wedge 
                          \ev_{[6,7]} \left( s \in P_3 \wedge 
                               \ev_{[8,9]} s \in P_4  
                          \right)   
                     \right) 
                 \right)
\label{eq:phi6} \\
\varphi_7 & = & \ev_{[5,8]} \left( s \in P_1 \wedge  
                     \ev_{[6,11]}\left(s \in P_2 \wedge 
                          \ev_{[6,7]} \left( s \in P_3 \wedge 
                               \ev_{\textcolor{red}{[9,10]}} s \in P_4  
                          \right)   
                     \right) 
                 \right)
\label{eq:phi7} 
\end{eqnarray}}


\myipara{Practical Exponential Stability} We next consider a linear
plant model (Eq.~\eqref{eq:expstabplant}) and a $\relu$-NN controller
that tries to stabilize the system to satisfy a practical exponential
stability criterion as expressed by the STL formula $\varphi_8$ in
\eqref{eq:expstab}; note that in $\varphi_8$,
$P_{6} \subset P_{5} \subset  \cdots \subset P_2 \subset P_1$.
\begin{equation}
\label{eq:expstabplant}
\statee_{\timeid+1} = A \statee_{\timeid} + B u(\statee_\timeid),\ A=\begin{bmatrix} 0.9105 & -0.9718\\  0.5177 & 0.3552 \end{bmatrix}, \ B= \begin{bmatrix} 0.21  &  0.05 \\ 0.15  &  -0.28 \end{bmatrix}.
\end{equation}
\begin{equation}
\begin{array}{ll}
\varphi_8 = & \alw_{[9,16]}\left[ s \in P_1\right] \wedge \alw_{[17,24]}\left[ s \in P_2\right] \wedge \alw_{[25,32]}\left[ s
\in P_3\right] \wedge \\
& \alw_{[33,40]}\left[ s \in P_4\right] \wedge  \alw_{[41,43]}\left[ s \in P_5\right] \wedge \alw_{[44,60]}\left[ s \in P_6\right]
\end{array}
\label{eq:expstab}
\end{equation}
The architecture of  NN controller is $\left[ 2,\ 30,\ 30,\ 30,\ 2 \right]$.
We attempt to verify if the controller satisfies the mentioned STL
specification for the inital state set $\mathcal{I}
= \left\{ (x,y) | x\in[-50,-40],\ y\in[85,95] \right\}$.  The regions $P_5$ and $P_6$ are small. This requires us to apply exact-star technique. On the other hand the exact-star is time consuming on $\init$ but partitioning $\init$ in $25$ partitions is quite helpful to verify within a reasonable running time.
The results are presented in Table \ref{tbl:expstb2}.
\begin{figure}
    \includegraphics[width=\linewidth]{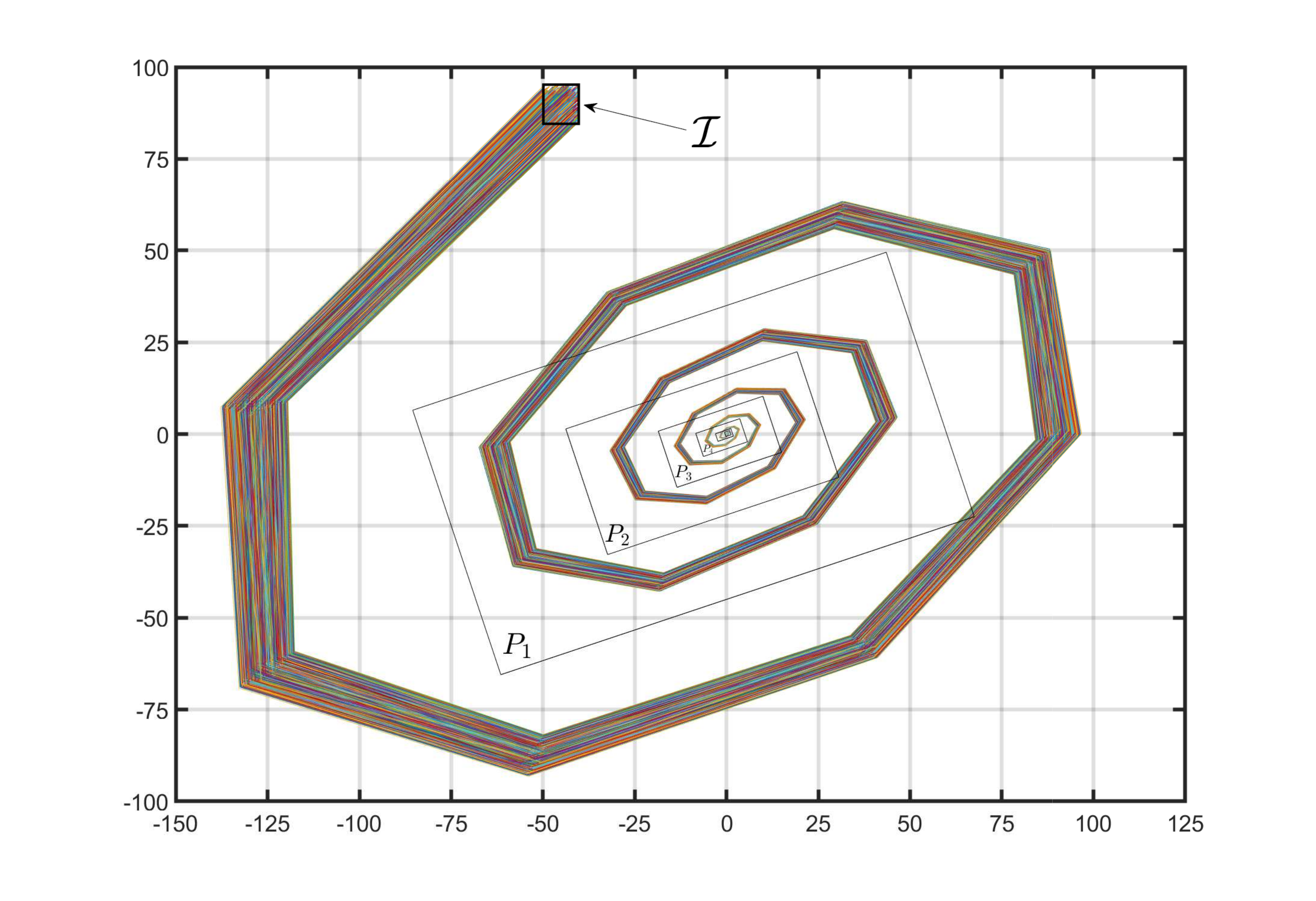}
    \captionof{figure}{\small{
		Trajectories for NNCS shown in
                Eq.~\eqref{eq:expstabplant}. 
		}
    } \label{fig:expstab2}
\end{figure}

\begin{table*}[t]
\centering
\resizebox{\textwidth}{!}{
\begin{tabular}{cccc|cccc} 
\hline 
$\init$& Robustness Range& Run-time & Verified? & $\init$& Robustness Range& Run-time & Verified? \\ \hline 
$x \in[-50,-48], y\in[85,87]$ & $\left[ 0.2380,\ 0.3173 \right]$ & 3199.4 $\mathrm{sec}$ & Yes \  & $x \in[-50,-48], y\in[87,89]$ & $\left[ 0.2414,\ 0.2884 \right]$ & 28.22 $\mathrm{sec}$ &Yes\\  
$x \in[-50,-48], y\in[89,91]$ & $\left[ 0.2383,\ 0.2638  \right]$ & 13.19 $\mathrm{sec}$ & Yes & $x \in[-50,-48], y\in[91,93]$ & $\left[ 0.2151,\ 0.2429 \right]$ & 114.3 $\mathrm{sec}$ & Yes\\  
$x \in[-50,-48], y\in[93,95]$& $\left[ 0.1927,\ 0.2244  \right]$ & 199.8 $\mathrm{sec}$ & Yes & $x \in[-48,-46], y\in[85,87]$ & $\left[ 0.2539,\ 0.3130 \right]$ &  287.3 $\mathrm{sec}$ & Yes\\  
$x \in[-48,-46], y\in[87,89]$ & $\left[ 0.2435,\ 0.2881 \right]$ & 2708.5 $\mathrm{sec}$ & Yes & $x \in[-48,-46], y\in[89,91]$ & $\left[ 0.2376,\ 0.2669 \right]$ & 2645.6 $\mathrm{sec}$ & Yes \\  
$x \in[-48,-46], y\in[91,93]$& $\left[ 0.2183,\ 0.2468 \right]$ & 45.8 $\mathrm{sec}$ & Yes \  & $x \in[-48,-46], y\in[93,95]$ & $\left[ 0.1934,\ 0.2228 \right]$ & 6.8 $\mathrm{sec}$ &Yes\\  
$x \in[-46,-44], y\in[85,87]$ & $\left[ 0.2824,\ 0.3140 \right]$ & 467.9 $\mathrm{sec}$ & Yes & $x \in[-46,-44], y\in[87,89]$ & $\left[ 0.2550,\ 0.2916  \right]$ & 1230.9 $\mathrm{sec}$  & Yes\\  
$x \in[-46,-44], y\in[89,91]$& $\left[ 0.2386,\ 0.2680  \right]$ & 1408.4 $\mathrm{sec}$& Yes & $x \in[-46,-44], y\in[91,93]$& $\left[ 0.2138,\ 0.2432  \right]$ &  610.1$\mathrm{sec}$ & Yes \\ 
$x \in[-46,-44], y\in[93,95]$ & $\left[ 0.1889,\ 0.2183 \right]$ & 16.7 $\mathrm{sec}$ & Yes & $x \in[-44,-42], y\in[85,87]$ & $\left[ 0.2839,\ 0.3133 \right]$ & 7.9 $\mathrm{sec}$ & Yes \\  
$x \in[-44,-42], y\in[87,89]$ & $\left[ 0.2590,\ 0.2884 \right]$ & 36.4 $\mathrm{sec}$ & Yes & $x \in[-44,-42], y\in[89,91]$& $\left[ 0.2341,\ 0.2635 \right]$ & 152.2 $\mathrm{sec}$ & Yes \\  
$x \in[-44,-42], y\in[91,93]$ & $\left[ 0.2092,\ 0.2386 \right]$ & 796.4 $\mathrm{sec}$ &Yes & $x \in[-44,-42], y\in[93,95]$ & $\left[ 0.1844,\ 0.2138 \right]$ & 1282.8 $\mathrm{sec}$ & Yes \\  
$x \in[-42,-40], y\in[85,87]$ & $\left[ 0.2793,\ 0.3087  \right]$ & 7.6 $\mathrm{sec}$  & Yes & $x \in[-42,-40], y\in[87,89]$& $\left[ 0.2545,\ 0.2839  \right]$ & 5.8 $\mathrm{sec}$ & Yes \\  
$x \in[-42,-40], y\in[89,91]$ & $\left[ 0.2296,\ 0.2590  \right]$ & $6 \mathrm{sec}$ & Yes & $x \in[-42,-40], y\in[91,93]$ & $\left[ 0.2047,\ 0.2341 \right]$ & $45.8 \mathrm{sec}$ & Yes \\  
$x \in[-42,-40], y\in[93,95]$ & $\left[ 0.1798,\ 0.2092 \right]$ & $142.4 \mathrm{sec}$ & Yes& --- & --- & --- & ---\\ \hline 
\end{tabular} 
}
\caption{Verifying $\varphi_8$ against NNCS in
Eq.~\eqref{eq:expstabplant} utilizing exact-star reachability on TNN. Initial state set $\init = \left\{ (x,y) | x\in[-50,-40],\ y\in[85,95] \right\}$. 
The trajectory encoding has $180$ layers and $\stltonn$ has $8$ layers. No parallel computing is utilized.} \label{tbl:expstb2}
\end{table*} 

\section{STL Verification using Sampling}
\label{sec:lip}

Consider the TNN structure described in Sec.~\ref{sec:Ver}. If we
compute the local Lipschitz constant of the function $\trpzmffnn$
w.r.t. the initial state $\statee_0$, then we can use this to obtain a
certificate that all initial states satisfy the given STL formula. The
basic idea is that if we sample the set of initial states {\em dense
enough}, and the value of $\trpzmffnn$ is {\em positive enough} at all
sample points, then this lets us reach a sound conclusion that
$\trpzmffnn$ is positive for all initial states. This intuition is
formalized in Theorem~\ref{thm:Lipbound}.

\begin{theorem}\label{thm:Lipbound}
Assume $L_{loc}$ is the local Lipstchiz constant of function $f: \mathbb{R}^n \to \mathbb{R}$ on the domain $[\ell
,u]$ where $\ell,u\in \mathbb{R}^n$ . We denote the set of all $2^n$ vertices on $[\ell,u]$ by
$V([\ell,u])$ and we assume,
$$
\forall x\in V([\ell, u])\ :\ f(x)> 0
$$
Given the certificates $\rho_1 > L_{loc}$ and $\rho_2=\underset{x \in V([\ell,u])}{\min} f(x)$,
$$
\|u-\ell\|_2 < \frac{\rho_2}{\rho_1} \ \implies \forall x \in [\ell, u] \ :\ f(x)> 0
$$
\end{theorem}

\proof Consider $x \in [\ell, u]$ and $x^* \in V([\ell,u]),\ f(x^*)=\rho_2$, this implies, $\|x^*-x\|_2 \leq \|u-\ell\|_2$. We know $\rho_1$ is an upper bound for the local Lipschitz constant $L_{loc}$, therefore,
$$
\resizebox{\hsize}{!}{$
\|\rho_2-f(x)\|_2 \leq \rho_1 \|x^*-x\|_2 \leq \rho_1\|u-\ell\|_2 \implies \|u-\ell\|_2 \geq \frac{\|\rho_2-f(x)\|_2}{\rho_1} $}
$$ 
We will prove by contradiction that $f(x)>0$. Assume $f(x)\leq 0$. Since $\rho_2>0$, we can conclude $\|u-\ell\|_2 \geq \rho_2/\rho_1$ which contradicts our assumption. \hspace*{\fill}~$\square$\\

The certificate $\rho_1$, is an upper bound for the local Lipschitz
constant of $\trpzmffnn(\statee_0)$ with respect to the initial state,
$\statee_0 \in \init$. If a bounded certificate $\rho_1$ is accessible
then we can utilize Theorem \ref{thm:Lipbound} for a sound and
complete verification of controllers. Based on Theorem
\ref{thm:Lipbound} we are required to select an $\epsilon>0$ to build
an $\epsilon\!$-net over the set of initial states. For every single
hypercube in the $\epsilon$-net we compute $\rho_2$ and check whether
$\epsilon< \rho_2/\rho_1$. In case this condition doesn't hold we
create a finer grid on the mentioned hypercube. We terminate the
process, return the counter example and reject the controller if we
face $\rho_2<0$. Otherwise, we continue until $\epsilon<
\rho_2/\rho_1$ for every single hypercube and verify the controller.

The efficiency of this technique is highly related to the tightness of
the upper-bound $\rho_1$. For instance, if the upper bound is large,
to obtain a verification result, $\epsilon$ tends to be very small,
greatly increasing the points over which to check the required
condition.  Thus, the key problem here is to solve the local Lipschitz
constant computation for neural networks. This problem has been
addressed by a variety of techniques in the literature
\cite{raghunathan2018semidefinite}, \cite{avant2020analytical}, \cite{latorre2020lipschitz}, \cite{jordan2020exactly} but there is limitation on
their time and memory scalability. The existent techniques in the
literature are mostly limited to $\relu$ activation functions. There
is also a trade-off between their scalability and accuracy. 

In this paper, we use the convex programming technique presented in
\cite{fazlyab2019efficient}, \cite{hashemi2021certifying} as convex
programming scales to larger neural networks with low conservatism.
The proposed technique
\cite{fazlyab2019efficient}, \cite{hashemi2021certifying} in its current
formulation is not directly applicable to our verification process but
we can apply it with small modifications; the details are discussed in
Appendix \ref{apdx:Lip-SDP}. We call this specific formulation of
proposed convex programming in \cite{hashemi2021certifying}, \cite{fazlyab2019efficient} as
$\mathsf{Trapezium\!\!-\!\!Lip\!\!-\!\!SDP}()$. We remark that this
method is applicable to plant and controller models that are neural
networks with {\em arbitrary} activation functions or plants that have
linear models. However, in its current form, we were not able to get
conclusive verification results for arbitrary nonlinear ODE-based
models (as local Lipschitz computation returned overly conservative
Lipschitz constant values).

\SetKwProg{Fn}{Function}{}{end}
\SetKwSwitch{Switch}{Case}{Other}{switch}{}{case}{otherwise}{endcase}{endsw}
\SetFuncSty{mysty}
\SetKw{Return}{return}
\SetKw{Break}{terminate;}
\DontPrintSemicolon
\begin{algorithm}[t]
\SetKwFunction{lipver}{Lip-Verify}
\SetKwFunction{stltocbf}{stl2cbf}
\Fn{\lipver{$\varphi,\rho_1,\init,\mathrm{model}, \mathrm{controller}, N, \mathsf{status}$}}{
           $-$ construct a uniform $\epsilon$-net of $N$ hypercubes
           over $\init$ \\
           $\epsilon\text{-net}
           =\overset{N}{\underset{i=1}{\bigcup}} [ \ell_i,\ u_i ]$, \quad
           $\epsilon_i=\| u_i-\ell_i\|_2$ \\
             \While{true}{
                \For{$i \gets 1$ to $N$}{
                    \If{$\mathsf{status} \neq \mathsf{Solved}$}{
                    $\rho_1, \mathsf{status} \gets \mathsf{Trapezium\!\!-\!\!Lip\!\!-\!\!SDP}([ \ell_i,\ u_i ])$}\;
                    $\rho_2 \gets \underset{x \in V([ \ell_i,\ u_i ])}{\min} \trpzmffnn(x)$ \;
                    \If{$\rho_2<0$}{
                        \Return Falsified + counter example\\
                        \Break 
                    }
                    \Else{
                        \If{$\left(\mathsf{status\neq \mathsf{Solved}}\right) \vee \left(\left(\mathsf{status} = \mathsf{Solved}\right) \wedge \left(\epsilon_i > \rho_2/\rho_1\right)\right)$}{
                        \Return \lipver{$\varphi,\rho_1,[ \ell_i,\ u_i ], \mathrm{model},\mathrm{controller},N, \mathsf{status}$ }
                       }
                   }
                }
                \Return Verified
            }
}
\caption{Recursive algorithm for verification with local Lipschitz certificates.}
\label{algo:Lipver}
\end{algorithm}
\subsection{Experimental Validation}\label{subsec:lipverifyresults}
We now present results of applying our Lipschitz constant
computation-based technique for verification.

\myipara{Simple $\mathtt{tanh}$-activation model} In this case study,
we consider plant and controller models with structure $[4,\  5,\
2]$, and $[2,\ 5, \ 2]$ respectively, where both models use the
hyperbolic tangent activation function. In this problem we verify the
STL formula shown in \eqref{eq:phi10}.Here, the specified set of initial states is provided as $\mathcal{I}=[1,2]\times[1,2]$.
\begin{equation}
\label{eq:phi10}
\varphi_{10} = \ev_{[3,6]} \left(\  [s \in P_1] \ \wedge \ \ev_{[9,13]} \ [s \in P_2] \right).
\end{equation}
The TNN model contains a total of $38$ hidden tangent hyperbolic (+
linear) layers for encoding the trajectory and 10 hidden $\relu$ (+ linear)
layers for $\stltonn$. We
first partition $\init$ into 4 squares (see Figure
\ref{fig:Lip_Verify}) where $\epsilon=\sqrt{2}/2$ for each set. We
employ the CROWN library \cite{wang2018efficient} for the
pre-activation bound computation on each trajectory layer. We also
utilize the approx-star technique \cite{tran2019star} for pre-activation
bound computation on the $\stltonn$. Then we utilize convex
programming approach $\mathsf{Trapezium\!\!-\!\!Lip\!\!-\!\!SDP}()$)
that we developed
with MOSEK \cite{andersen2000mosek} and YALMIP
\cite{lofberg2004yalmip} solvers to compute $\rho_1$. We also utilize
$\stltonn$ for each partition to compute the certificate $\rho_2$. The
results are shown in Figure \ref{fig:Lip_Verify}.  
In the first round of partitioning, the
desired condition $\epsilon \leq \rho_2/\rho_1 $ does not hold for any
partition. This implies we must partition all $4$ subsets
(see Figure \ref{fig:Lip_Verify}). In the next round of partitioning,
$\epsilon=\sqrt{2}/4$ and $8$ subset from $16$ are verified satisfying
$(\epsilon \leq \rho_2/\rho_1)$. For the remaining $8$ non-verified
subsets we apply the third round of partitioning resulting in
$\epsilon=\sqrt{2}/8$ where all of them become verified. Figure
\ref{algo:Lipver} presents the flow of recursive algorithm \ref{algo:Lipver} With $3$ recursive calls. The verification concludes after 90 seconds with this algorithm.

\begin{figure*}
  \begin{minipage}{0.55\textwidth}
  
  \includegraphics[width=\linewidth]{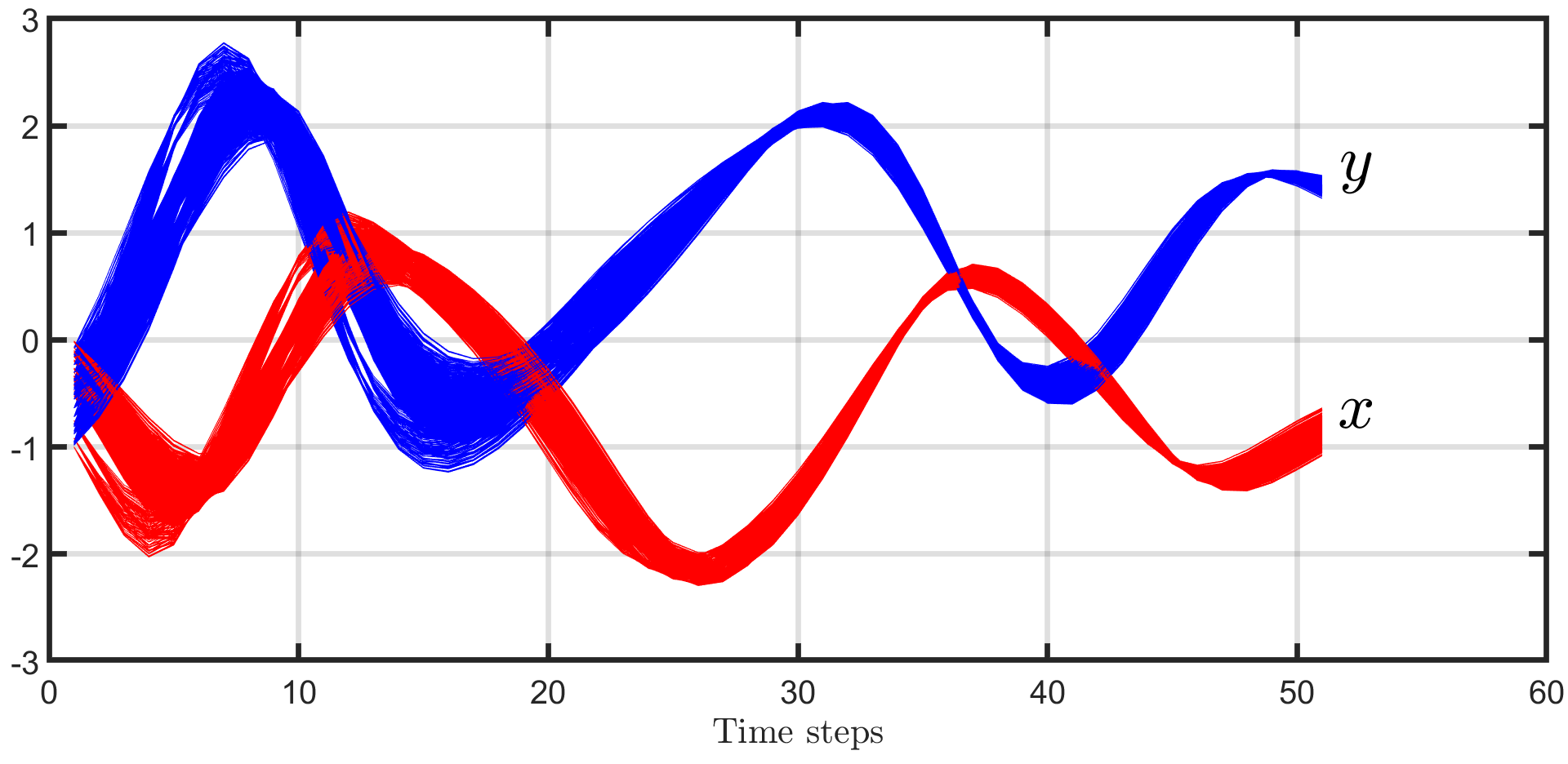}
  \caption{\small{Shows the evolution of states in a control feedback system for proposed LTV model in 50 time steps.}}\label{fig:LTV_reachset}
  \end{minipage}
  \begin{minipage}{0.41\textwidth}
  \includegraphics[width=\linewidth]{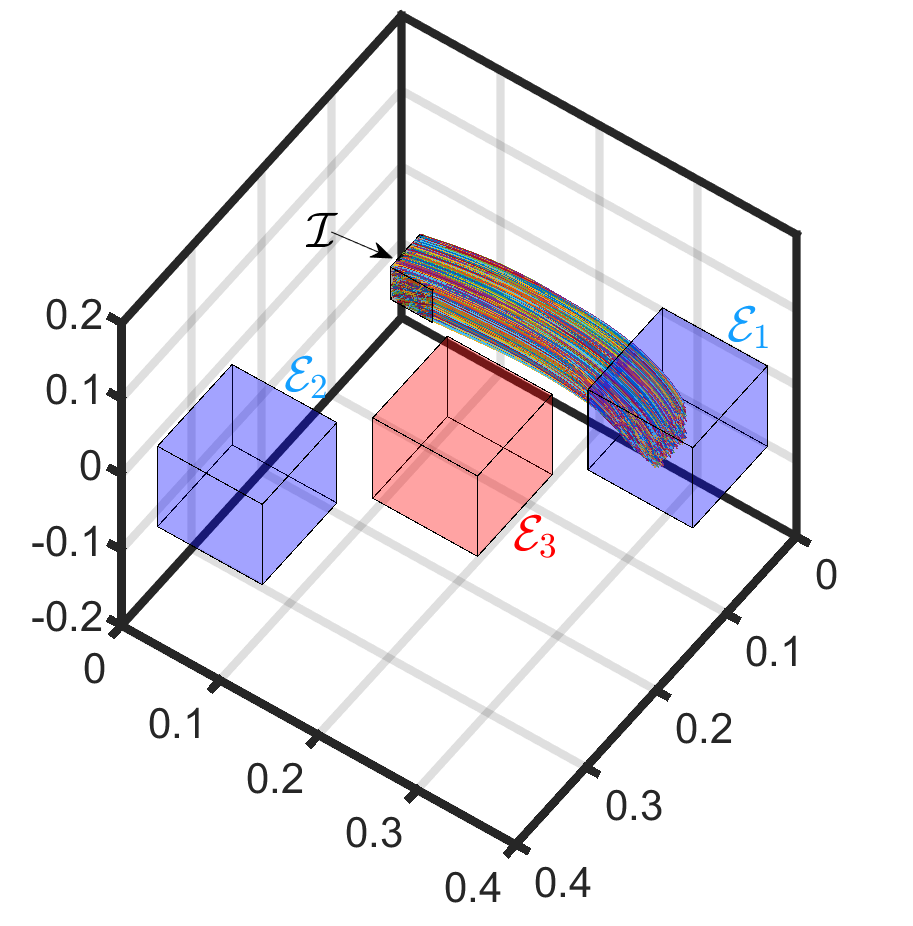}
  \caption{\small{Shows the evolution of states for the quadrotor example, the quadrotor is controlled with a pre-trained $\tanh$ FFNN controller, the quadrotor is planned to avoid $\mathcal{E}_3$ but requires to meet on of destinations $\mathcal{E}_1$ or $\mathcal{E}_2$ within $20$ time steps.}}\label{fig:quad_reachset}
\end{minipage}
\end{figure*}

\myipara{Linear Time-Varying Plant}
Figure \ref{fig:LTV_reachset} shows the evolution of control feedback
system with the following LTV model,
where,
$$
A(\tau)=\begin{bmatrix}0 & 1 \\ -2-sin(\tau) & -1 \end{bmatrix},\  B=\begin{bmatrix}1\\0\end{bmatrix},\ T=\frac{2\pi}{30},\  \statee_\timeid = \begin{bmatrix}x_\timeid\\ y_\timeid\end{bmatrix}
$$
which is the Zero-Order Hold discretization of
$\dot{\statee}=A(t)\statee + B \NN(\statee)$, with sampling time $T$.
The controller is a neural network of structure $[2,7,7,1]$, with
$\tanh()$ activation function and is expected to satisfy,
$$
\begin{aligned}
\varphi &= \ev_{[30,35]}\left[x_\timeid \leq -0.5\right] \bigwedge \ev_{[37,43]}\left[x_\timeid \geq -0.4\right] \bigwedge \ev_{[45,50]}\left[x_\timeid \leq 0\right]\\
&\bigwedge \ev_{[32,38]}\left[y_\timeid \geq 1\right] \bigwedge \alw_{[1,50]}\left[y_\timeid-x_\timeid \leq 5.5\right].
\end{aligned}
$$
Since the parallel computing does not support recursive algorithms, we manually partition $\init:=\left\{\statee_0 \mid
\begin{bmatrix}-1,-1\end{bmatrix}^\top \leq \statee_0 \leq
\begin{bmatrix}0,0\end{bmatrix}^\top \right\}$ into 64 equal subsets,
and run Algorithm~\ref{algo:Lipver} on every set. The average verification time for each sub-problem was around $6$ minutes. See Figure \ref{fig:LTV_runtime} for more detail.

\myipara{Neural Network Controlled Quadrotor System}
Figure~\ref{fig:quad_reachset} shows the evolution of control feedback
system  for a quadrotor. The model is trained on the following
dynamics with $T=0.05$ and trajectories start from $\init$,
$$
\begin{bmatrix} \dot{p}_x \\ \dot{p}_y \\ \dot{p}_z \\ \dot{v}_x \\ \dot{v}_y \\ \dot{v}_z \end{bmatrix} = \begin{bmatrix}v_x\\ v_y \\ v_z \\ \mathrm{g}\tan(\theta)\\ -\mathrm{g} \tan(\phi) \\ \tau-\mathrm{g} \end{bmatrix},\  \init=\left\{  \statee_0\ \mid \  \begin{bmatrix} 0.0638\\0.0638 \\ -0.0213 \\ 0 \\ 0 \\ 0\end{bmatrix} \leq \statee_0 \leq \begin{bmatrix} 0.1063\\0.1063 \\ 0.0213 \\0 \\ 0 \\ 0\end{bmatrix} \right\}
$$
We train a $\tanh()$ FFNN on this dynamics using $3.696 \times 10^6$ training data. The model 's dimension is $[9,10,10,6]$. The controller is also $\tanh()$ FFNN with dimension $[6,10,3,3]$. We wish to verify the formula:
$$
\varphi=\ev_{[1,20]} \left( \left[s\in \mathcal{E}_1\right] \vee \left[ s \in \mathcal{E}_2 \right] \right) \bigwedge \alw_{[1,20]} \left[ s \notin \mathcal{E}_3 \right]
$$
Here the controller is time-varying and its first bias vector
linearly varies with time. ($b_1(\timeid)=\bar{b}_1+\timeid \delta
b_1$). Since the parallel computing does not support recursive algorithms,  we manually partition $\init$ into $64$ equal cubes and run the algorithm
\ref{algo:Lipver} on every one of them. The approximate running time for
the majority of them was $40$ minutes. But for some regions the verification was time consuming. See Figure \ref{fig:Q_runtime} for more detail.

\begin{table*}
\begin{subfigure}[a]{\textwidth}
  \centering
  \includegraphics[width=\linewidth]{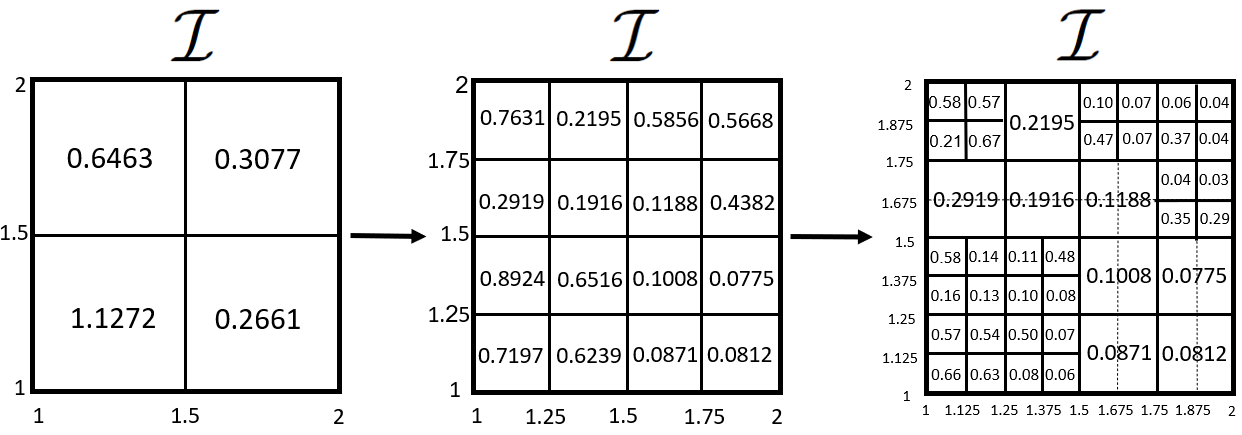}
  \caption{}
\end{subfigure}

\begin{subfigure}[b]{\textwidth}
  \centering
  \includegraphics[width=\linewidth]{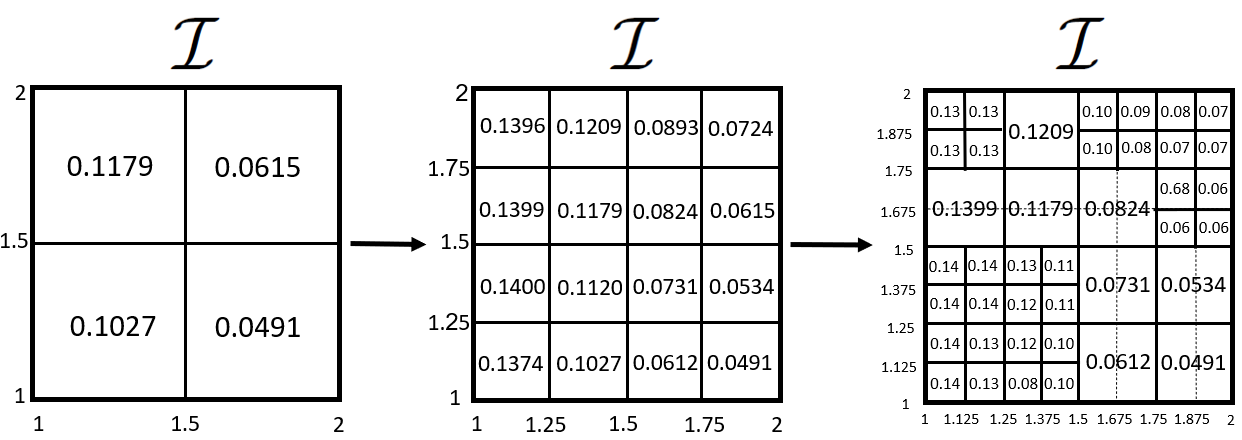}
  \caption{}
\end{subfigure}

\begin{subfigure}[c]{\textwidth}
  \centering
  \includegraphics[width=\linewidth]{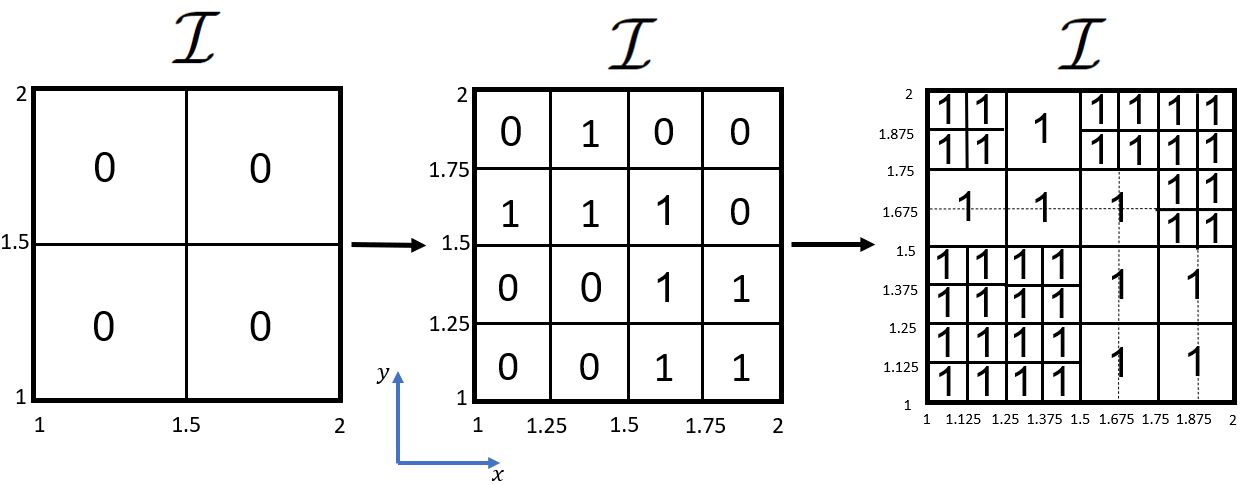}
  \caption{}
\end{subfigure}
\vspace{-4mm}
\captionof{figure}{\small{[Recursive partitioning for STL verification
with local Lipschitz computation: (a) Presents the certificates $\rho_1$ for each partition at each step. These certificates are computed with convex programming utilizing MOSEK and YALMIP. The results are rounded upwards. (b) Presents the certificates $\rho_2$ for each partition at each step. These certificates are computed over Trapezium-FFNN. The results are rounded downwards. (c) shows the verification results. The result is $0$ when $\epsilon> \rho_2/\rho_1$ and is $1$ when $\epsilon < \rho_2/\rho_1$. Obviously $1$ indicates the controller is verified over the subset. We partitioned $\init$ in three steps to receive $1$ on every partition. The diameter $\epsilon$ is $\sqrt{2}/2,\ \sqrt{2}/4,\ \sqrt{2}/8$ for the biggest, medium and smallest partitions respectively.}}
\label{fig:Lip_Verify}
\end{table*}



\section{Related Work \& Conclusions}
\label{sec:conclusion}
\myipara{Related work} Safety verification of NNCS is well studied in the literature. We can classify these works in two categories. One group addresses open loop control systems.   The authors in \cite{katz2017reluplex} present a verification technique based on Satisfiability Modulo Theories (SMT). They extend simplex to handle $\relu$ activation functions and propose an efficient verification for $\relu$ networks called ReLUplex. They also propose in \cite{katz2019marabou} another technique based on SMT called Marabou, which is not restricted on $\relu$ activation function. The authors in \cite{pulina2010abstraction} propose verification for multi-layer perceptrons using abstraction to Boolean combination of linear arithmetic constraints which is also based on SMT. The authors in \cite{dutta2017output}, \cite{kouvaros2018formal}, \cite{lomuscio2017approach} present several verification techniques that are based on Mixed Integer Linear Programming MILP and the works \cite{anderson2019optimization}, \cite{gehr2018ai2}, \cite{singh2018fast}, \cite{singh2019abstract}, \cite{wang2018formal}, \cite{wang2018efficient}, \cite{xiang2018output} are also considered as set based verification techniques. The other group addresses the closed loop NNCS. The authors in \cite{sun2019formal} propose sound and complete verification for discrete plants based on Satisfiability Modulo Convex (SMC) techniques. The authors in \cite{dutta2019reachability} propose a fast and efficient algorithm that is restricted on $\relu$ activation function based on regressive polynomial rule inference. For Verification on NNCS with ODE models the authors in \cite{huang2019reachnn} propose a reachability analysis on nonlinear plants employing Taylor series and Bernstein Polynomials. This method is not restricted on $\relu$ and is adjustable to control the level of conservatism.  Falsification and test-based approaches are also introduced to the verification community with authors in \cite{dreossi2019compositional,dreossi2019verifai,tuncali2018simulation}.

\myipara{Conclusion} 
We present $\stltonn$ a $\relu$ network which can be utilized for neural network verification with general STL specifications over discrete time signals. Since the formulation of verification problem is highly dependent to the STL specifications we present Trapezium-FFNN as a basic structure for problem formulation which is quite helpful to provide a toolbox on this approach. This is the first toolbox for sound and complete verification with general STL specifications. Unlike the other verification toolboxes, this toolbox is not restricted to $\relu$ networks. For a neural network controlled system with difference equation models, our $\stltonn$ can be applied on existing approximate reachability techniques such as ReachNN  \cite{huang2019reachnn} to provide a sound but not complete verification. 


\section*{Acknowledgement}
The authors would like to thank the anonymous reviewers for their feedback. This work was supported by the National Science Foundation through the following grants: CAREER award $\mathsf{SHF-2048094}$, $\mathsf{CNS-1932620}$, and funding by Toyota R\&D through the USC Center for Autonomy and AI.

\bibliographystyle{splncs04}
\bibliography{main}

\appendix

\section{Appendix}
\label{sec:appendices}
\subsection{Logarithmic Extension for Lemma \ref{lem:reluminmax}} \label{apdx:generalminmax}

\begin{figure}
\centering
    \includegraphics[width=0.5\linewidth]{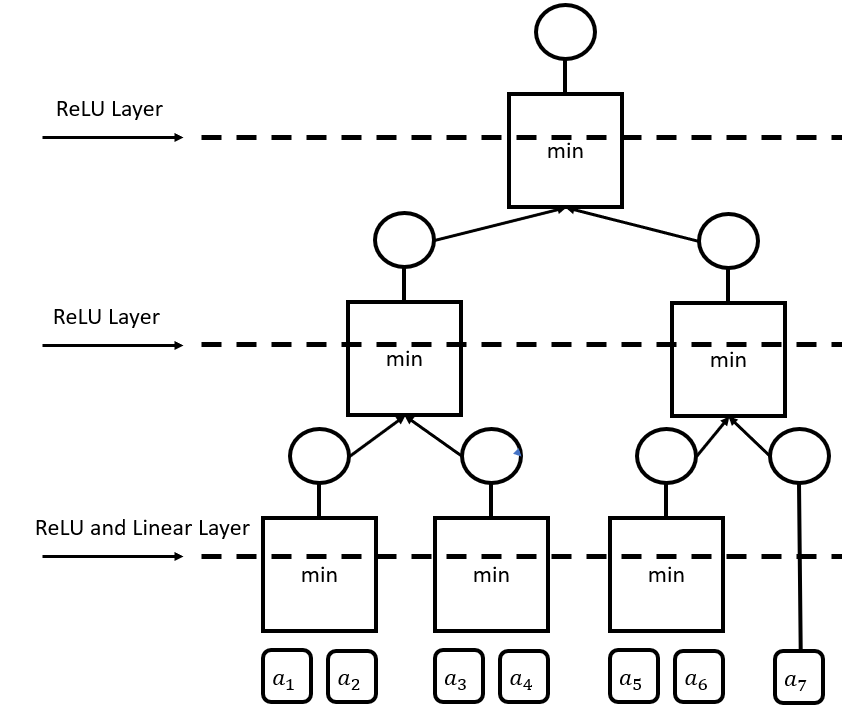}
    \caption{\small{Shows the structure to find the minimum of 7 variables.
    This structure provides us a FFNN, for the computation of min
    value. The depth of FFNN increases logarithmically with the number
    of inputs. In this approach we split the set of inputs in the
    pairs of two. Then we apply min ReLU network on each pair and
    continue this recursive algorithm until the minimum is achieved.
    The presence of the elements like $a_7$, which is not involved in
    min computation at the first layer, requires us to introduce
    linear activation functions in companion with $\relu$ in that
    layer, since we need to construct a unique FFNN that receives all
    the 7 elements and returns the minimum of them.} }
	\label{fig:poslin-maxmin}
\end{figure}

We are interested in an extension for Lemma \ref{lem:reluminmax} that
provides a network with depth of logarithmic order corresponding to
the number of inputs. Thus we group the inputs in pairs of two and
apply Lemma \ref{lem:reluminmax} repeatedly. Figure \ref{fig:poslin-maxmin} clarifies
this extension for a set of $7$ inputs.

\subsection{Lipschitz Constant Analysis for TNN}\label{apdx:Lip-SDP}
Upper bound for local Lipschitz constant of a FFNN is derived in \cite{hashemi2021certifying}, \cite{fazlyab2019efficient}. The presence of linear activation functions in $\stltonn$ must not impose computational complexity but if we include them in the proposed procedure in \cite{hashemi2021certifying}, \cite{fazlyab2019efficient} the optimization process faces memory problems as the size of LMI increases unnecessarily. Thus we slightly modify the proposed solution. We call this slightly modified version as $\mathsf{Trapezium\!\!-\!\!Lip\!\!-\!\!SDP}()$. Here we propose a summary of the convex programming approach from \cite{hashemi2021certifying}, \cite{fazlyab2019efficient} including the slight changes we apply on it.

Let's define the SDP variable $\rho=\rho_1^2$. We can reformulate the Lipschitz inequality $\|f(x_1)-f(x_2)\|_2 \leq \sqrt{\rho} \|x_1-x_2\|_2$ in the form of linear quadratic constraint as follows:
$$
  \begin{bmatrix} x_1-x_2\\ f(x_1)-f(x_2)\end{bmatrix}^\top \begin{bmatrix} \rho I_n  & 0_{n\times1}\\ 0_{1 \times n} & -1 \end{bmatrix} \begin{bmatrix} x_1-x_2\\ f(x_1)-f(x_2) \end{bmatrix} \geq 0 
$$
and we can conclude if,
$$
\begin{bmatrix} \rho I_n  & 0_{n\times1}\\ 0_{1 \times n} & -1 \end{bmatrix} \geq 0 \ \text{(positive semi-definite)}
$$
then $\rho_1=\sqrt{\rho}$ is certainly the desired certificate. Unfortunately due to presence of the negative scalar $-1$, this constraint is infeasible and we attempt to provide feasibility with provision of new linear information about function $f$. Thus the basic idea of convex programming technique is to provide the best symmetric linear matrix $Q_{\mathrm{info}}$ and transformation matrix $T$ that bring feasibility for,
\begin{equation}\label{eq:feasibility}
Q_{\mathrm{info}}-T^\top \begin{bmatrix} \rho I_n  & 0_{n\times1}\\ 0_{1 \times n} & -1 \end{bmatrix} T \leq 0,
\end{equation}
where $Q_{\mathrm{info}}$ is a linear combination of quadratic
constraints (QC), where every single QC represents a linear
information about function $f$. In this constraint $\rho=\rho_1^2$,
where $\rho_1$ is the certificate introduced in Theorem \ref{thm:Lipbound} for  $f: [\ell,\  u] \to [a,\ b]$, $\ell,u \in \mathbb{R}^n$ and $a,b\in \mathbb{R}$.  We add new information utilizing s-procedure technique proposed in \cite{fazlyab2019efficient}, \cite{hashemi2021certifying}. A thorough introduction for computation of $Q_{\mathrm{info}}$ is provided in \cite{fazlyab2019efficient}, \cite{hashemi2021certifying}. Provision of high quality information results in feasibility and tightness, but the presence of insufficient information results in infeasibility.

\subsubsection{QC for Non-linearities in feed-back structure:}
Figure \ref{fig:TFBS} shows the layers of TNN. The layers of TNN are entitled with $n=1,\cdots,N$. These layers are departed into nonlinear and linear portions. The pre-activation of nonlinear portion, $p_\ell$ is fed in nonlinear portion and results in post-activation $z_\ell$.

Assume $\statee_0^1, \statee_0^2$ are two initial states. They provide the pre-activations $p_\ell^1, p_\ell^2 \in \mathbb{R}^{n_\ell}$ on the TNN. The post-activations are also $z_\ell^1, z_\ell^2 \in \mathbb{R}^{n_\ell}$ respectively. We denote $\delta p_\ell := p_\ell^1 -p_\ell^2$,  $\delta z_\ell := z_\ell^1 -z_\ell^2$. We inform the convex programming about nonlinearities in $\trpzmffnn(\statee_0)$ through the following quadratic constraints:\\
-\noindent \myipara{The nonlinearity is a vector of differentiable activation functions}   
    \begin{lemma}\label{lem:slprstqc} \cite{hashemi2021certifying}:
    Let $\phi(w) = (\sigma(w_1),\cdots,\sigma(w_n)), \ w \in \mathcal{X} \subseteq \mathbb{R}^m$, where $\sigma$ is differentiable. Define $e_i^\top \vec{\alpha} = \inf_{w \in \mathcal{X}} \ \sigma'(w_i)$ and $e_i^\top \vec{\beta} = \sup_{w \in \mathcal{X}} \ \sigma'(w_i)$. Then $\phi$ satisfies the  $\delta\mathrm{QC}$ defined by $(\mathcal{X},\mathcal{Q})$, where
    
    \begin{equation}
    \resizebox{\hsize}{!}{$
    \begin{aligned}
		\mathcal{Q} = \{Q \mid Q =  \begin{bmatrix}
			-2\operatorname{diag}(\vec{\alpha} \circ \vec{\beta} \circ \lambda ) & \operatorname{diag}((\vec{\alpha}+\vec{\beta}) \circ \lambda )  \\ \operatorname{diag}((\vec{\alpha}+\vec{\beta}) \circ \lambda )  & -2 \operatorname{diag}(\lambda)
		\end{bmatrix}, \ \lambda \in \mathbb{R}^m_{+} \}.
    \end{aligned}$}
    \end{equation}
    
    \end{lemma}
    \noindent Thus, we firstly compute vector of slope bounds through the pre-activation bound computation and as an example given the slope bounds  $\vec{\alpha}_\ell$ and $\vec{\beta}_\ell$ on the $\ell$-th layer of TNN we claim:
    $$
    \resizebox{\hsize}{!}{$
    \begin{bmatrix}\delta p_\ell\\\delta z_\ell\end{bmatrix}^\top \underbrace{\begin{bmatrix}
			-2\operatorname{diag}(\vec{\alpha}_\ell \circ \vec{\beta}_\ell \circ \lambda_\ell ) & \operatorname{diag}((\vec{\alpha}_\ell+\vec{\beta}_\ell) \circ \lambda_\ell )  \\ \operatorname{diag}((\vec{\alpha}_\ell+\vec{\beta}_\ell) \circ \lambda_\ell )  & -2 \operatorname{diag}(\lambda_\ell)
		\end{bmatrix}}_{Q_\ell}\begin{bmatrix}\delta p_\ell\\ \delta z_\ell \end{bmatrix} \geq 0 , \ \lambda_\ell \in \mathbb{R}^{n_\ell}_{+}\\
    $}
    $$
    -\myipara{The nonlinearity is a vector of non-differentiable activation functions}  
    \begin{lemma}\label{lem:slprstqcndiff}\cite{hashemi2021certifying}:
    Let $\phi(w)=\max(\alpha w,\beta w),  \ w \in \mathcal{X} \subseteq \mathbb{R}^m, 0 \leq \alpha \leq \beta <\infty$ and define $\mathcal{I}^{+}$, $\mathcal{I}^{-}$, and $\mathcal{I}^{\pm}$ as the set of activations that are known to be always active, always inactive, or unknown on $\mathcal{X}$, i.e., $\mathcal{I}^{+}= \{i  \mid w_i \geq 0,  \forall w \in \mathcal{X}\}$, $\mathcal{I}^{-}= \{i  \mid w_i < 0, \forall  w \in \mathcal{X}\}$, and $\mathcal{I}^{\pm}= \{1,\cdots,m\} \setminus (\mathcal{I}^{+} \cup \mathcal{I}^{-})$. Define $\balphha = [\alpha+(\beta-\alpha)\mathbf{1}_{\mathcal{I}^{+}}(1),\cdots,\alpha+(\beta-\alpha)\mathbf{1}_{\mathcal{I}^{+}}(m)]$ and $\bbetta = [\beta-(\beta-\alpha)\mathbf{1}_{\mathcal{I}^{-}}(1),\cdots,\beta-(\beta-\alpha)\mathbf{1}_{\mathcal{I}^{-}}(m)]$. Then $\phi$ satisfies the $\delta\mathrm{QC}$ defined by $(\mathcal{X},\mathcal{Q})$, where
    \begin{equation}
	\begin{aligned}
		\mathcal{Q} = \{Q \mid Q &=  \begin{bmatrix}
			-2\operatorname{diag}(\balphha \circ \bbetta \circ \lambda ) & \operatorname{diag}((\balphha+\bbetta) \circ \lambda )  \\ \operatorname{diag}((\balphha+\bbetta) \circ \lambda )  & -2\operatorname{diag}(\lambda)
		\end{bmatrix},\\
            &\ e_i^\top\lambda \in \mathbb{R}_{+} \ \text{ for }  i \in \mathcal{I}^{\pm}\}.
        \end{aligned}
    \end{equation}
    \end{lemma}
    \noindent Therefore, To capture the slope bounds, we firstly determine $\mathcal{I}^+,\mathcal{I}^-$ through the pre-activation bound computation and as an example given the slope bounds  $\balphha_\ell$ and $\bbetta_\ell$ on the $\ell$-th layer of TNN we claim for, $e_j^\top\lambda_\ell \in \mathbb{R}_{+} \ \text{ and } j \in \mathcal{I}^{\pm}$:
    $$
            \begin{bmatrix}\delta p_\ell \\ \delta z_\ell \end{bmatrix}^\top \underbrace{\begin{bmatrix}
			-2\operatorname{diag}(\balphha_\ell \circ \bbetta_\ell \circ \lambda_\ell ) & \operatorname{diag}((\balphha_\ell+\bbetta_\ell) \circ \lambda_\ell )  \\ \operatorname{diag}((\balphha_\ell+\bbetta_\ell) \circ \lambda_\ell )  & -2\operatorname{diag}(\lambda_\ell)
		\end{bmatrix}}_{Q_\ell}\begin{bmatrix} \delta p_\ell \\ \delta z_\ell \end{bmatrix} \geq 0,
    $$
  
\subsubsection{S-procedure for $Q_{\text{info}}$:} 
\begin{figure}
    \centering
    \vspace{-10mm}
    \includegraphics[width=\linewidth]{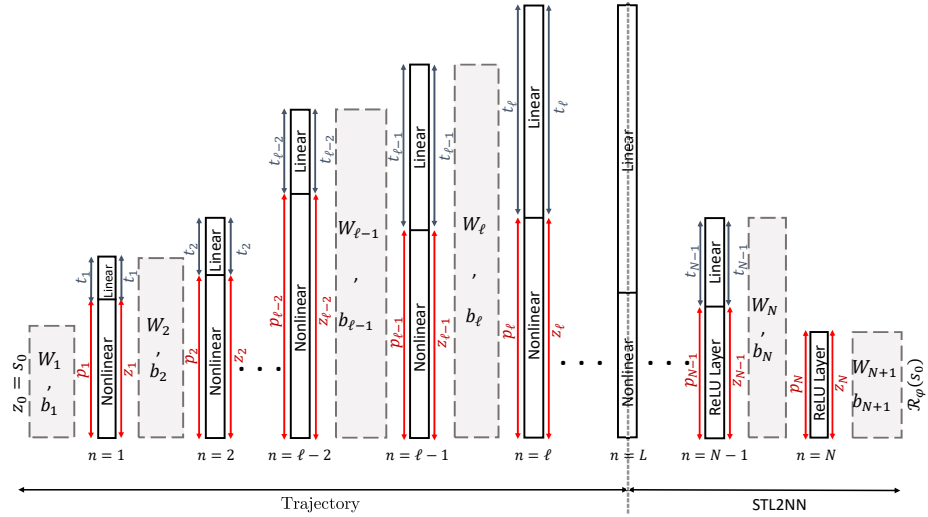}
    \vspace{-15mm}
    \caption{\small{Shows the TNN structure. Here $N$ is the number of layers on TNN. $[t_\ell,z_\ell]^\top$ presents the activation vector for $\ell$-th layer and $[t_\ell,p_\ell]^\top$ presents its pre-activation on TNN. The role of a linear activation function is to copy its input}}
    \label{fig:TFBS}
\end{figure}

Consider the weigh matrices on TNN in Figure \ref{fig:TFBS}. This weigh matrices are built from $4$ sub-blocks,
\begin{itemize}
    \item $W_\ell^{ll}$: This subblock connects the linear portion of layer $\ell-1$ to linear portion of layer $\ell$.
    \item $W_\ell^{ln}$: This subblock connects the linear portion of layer $\ell-1$ to non-linear portion of layer $\ell$.
    \item $W_\ell^{nl}$: This subblock connects the non-linear portion of layer $\ell-1$ to linear portion of layer $\ell$.
    \item $W_\ell^{nn}$: This subblock connects the non-linear portion of layer $\ell-1$ to non-linear portion of layer $\ell$.
\end{itemize}
 Therefore the pre-activation $p_\ell$ is computed as a linear combination of previous activations $z_i, i=0,\cdots, \ell-1$ through the following iterative formula.
$$
    \begin{bmatrix} t_1 \\ p_1 \end{bmatrix} = \underbrace{\begin{bmatrix} 
     W_1^{ln}\\
    \\ \hline \\ W_1^{nn} 
     \end{bmatrix}}_{W_1} \underbrace{z_0}_{\statee_0}+ b_1,  \ \begin{bmatrix} t_\ell \\\\ p_\ell \end{bmatrix} = \underbrace{\left[\begin{matrix}  
    \begin{array}{c|c}  
     W_\ell^{ll} \hspace{2mm}&  \hspace{2mm} W_\ell^{ln}\\
    \\ \hline \\
     W_\ell^{nl} \hspace{2mm}&  \hspace{2mm}   W_\ell^{nn} \end{array}  
     \end{matrix}\right]}_{W_\ell,\ \  \ell=2,\cdots,N} \begin{bmatrix}
     t_{\ell-1}\\\\ {z_{\ell-1}}\end{bmatrix}+b_{\ell},\  
$$
then the difference of pre-activations $\delta p_\ell$ are,
$$
\resizebox{\hsize}{!}{$
\begin{aligned}
& \delta p_1 =W_1^{nn}(\delta z_0), \quad \delta p_2 = W_2^{nl}W_1^{ln} \delta z_1 + W_2^{nn} \delta z_0\\
&\delta p_\ell = W_\ell^{nl} \sum_{k=0}^{\ell-3} \left[ \left( \prod_{j=1}^{\ell-k-2} W_{\ell-j}^{ll} \right)W_{k+1}^{ln}(\delta z_k)\right] +W_\ell^{nl}W_{\ell-1}^{ln}(\delta z_{\ell-2})+W_\ell^{nn} \delta z_{\ell-1}, \ \ell \geq 3 
\end{aligned}$}
$$
We also follow the process of \cite{hashemi2021certifying}, \cite{fazlyab2019efficient} and concatenate the non-linear activation vectors in the base vector $\delta Z:= [\delta z_0^\top, \cdots, \delta z_N^\top]^\top$. Given the proposed relation between $\delta p_\ell$ and $\delta Z$ we define the transformation matrix $E_\ell, T$ as,
$$
\begin{bmatrix}\delta p_\ell \\  \delta z_\ell \end{bmatrix}= E_\ell \  \delta Z, \ \ell=1,\cdots,N, \  
\begin{bmatrix}\delta z_0 \\  W_{N+1} \delta z_N\end{bmatrix} = T \ \delta Z 
$$
Finally based on the idea in \cite{hashemi2021certifying}, \cite{fazlyab2019efficient}, we claim if,
\begin{equation} \label{eq:result}
\sum_{\ell=1}^N \begin{bmatrix}\delta p_\ell \\  \delta z_\ell \end{bmatrix}^\top Q_\ell \begin{bmatrix}\delta p_\ell \\  \delta z_\ell \end{bmatrix} - \begin{bmatrix}\delta z_0 \\  W_{N+1} \delta z_N\end{bmatrix}^\top \begin{bmatrix} \rho I_n  & 0_{n\times1}\\ 0_{1 \times n} & -1 \end{bmatrix}\begin{bmatrix}\delta z_0 \\  W_{N+1} \delta z_N\end{bmatrix} \leq 0
\end{equation}
Then, $\norm{\trpzmffnn(\statee_0^1)-\trpzmffnn(\statee_0^2)}_2  \leq \sqrt{\rho} \norm{\statee_0^1-\statee_0^2}_2 $ and this is because equation \eqref{eq:result} implies,
$$
\begin{bmatrix}\delta z_0 \\  W_{N+1} \delta z_N\end{bmatrix}^\top \begin{bmatrix} \rho I_n  & 0_{n\times1}\\ 0_{1 \times n} & -1 \end{bmatrix}\begin{bmatrix}\delta z_0 \\  W_{N+1} \delta z_N\end{bmatrix} \geq 0
$$
This result  certifies $\rho_1=\sqrt{\rho}$ to be a true certificate as an upper bound of Lipschitz constant. On the other hand equation \eqref{eq:result} can be rephrased based on the base vector $\delta Z$ as,
\begin{equation} \label{eq:result2}
\delta Z^\top \left(  \sum_{\ell=1}^N E_\ell^\top Q_\ell E_\ell - T^\top \begin{bmatrix} \rho I_n  & 0_{n\times1}\\ 0_{1 \times n} & -1 \end{bmatrix}T \right) \delta Z \leq 0
\end{equation}
and proposing $Q_{\text{info}}= \sum_{\ell=1}^N E_\ell^\top Q_\ell E_\ell$ a sufficient condition to satisfy \eqref{eq:result2} is,
$$
Q_{\text{info}} - T^\top \begin{bmatrix} \rho I_n  & 0_{n\times1}\\ 0_{1 \times n} & -1 \end{bmatrix}T \leq 0
$$

\begin{figure}
    \includegraphics[width=\linewidth]{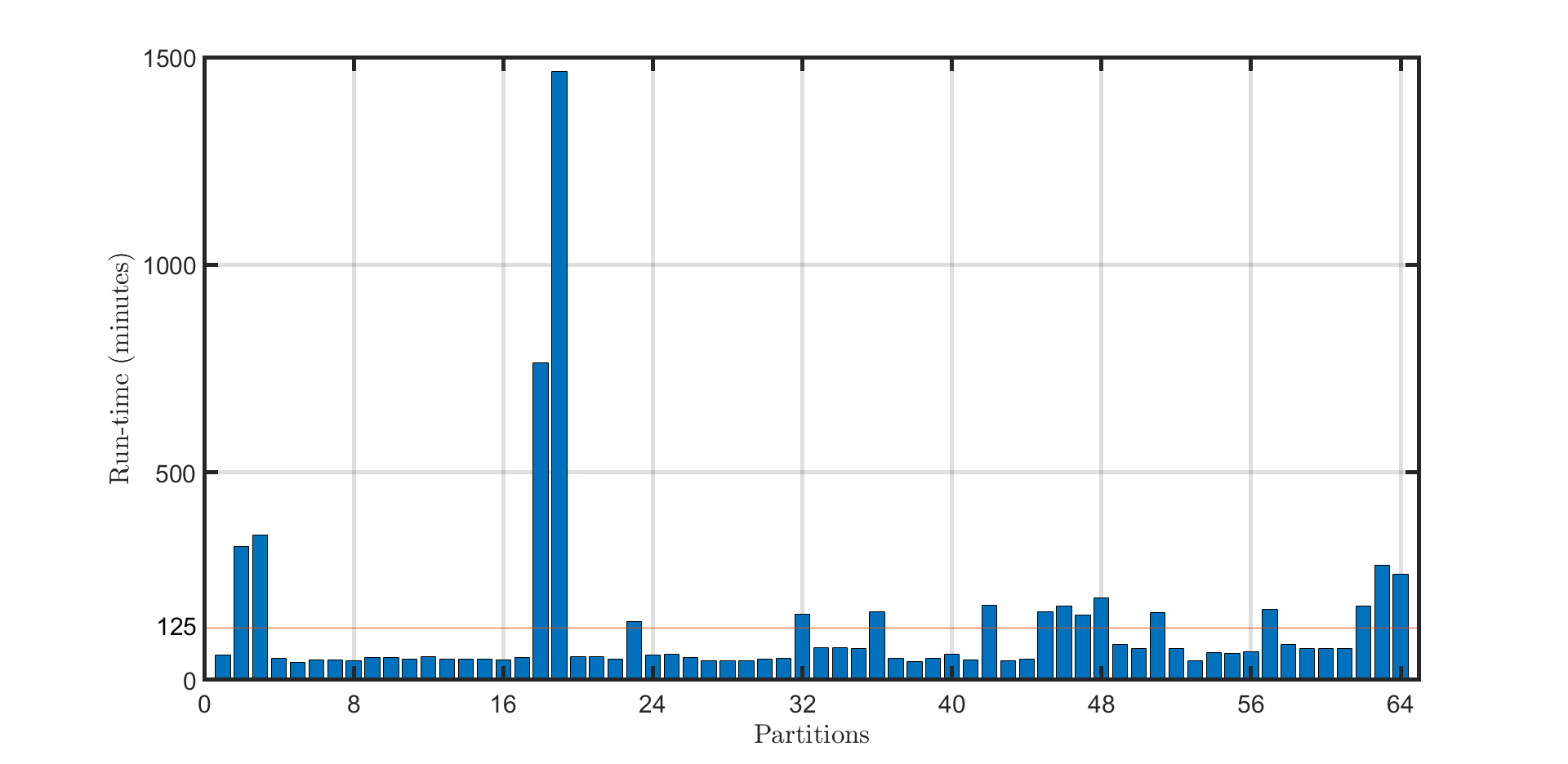}
    \caption{\small{Shows the verification run-time in Neural Network Controlled Quadrotor System, for every $64$ partitions of the set of initial states. We apply algorithm \ref{algo:Lipver} on every partition and conclude the verification. This figure shows the verification run time on the majority of partitions is approximately $40$ minutes. The red line shows the average run-time which is approximately $125$ minutes.}} \label{fig:Q_runtime}
\end{figure}
\begin{figure}
    \includegraphics[width=\linewidth]{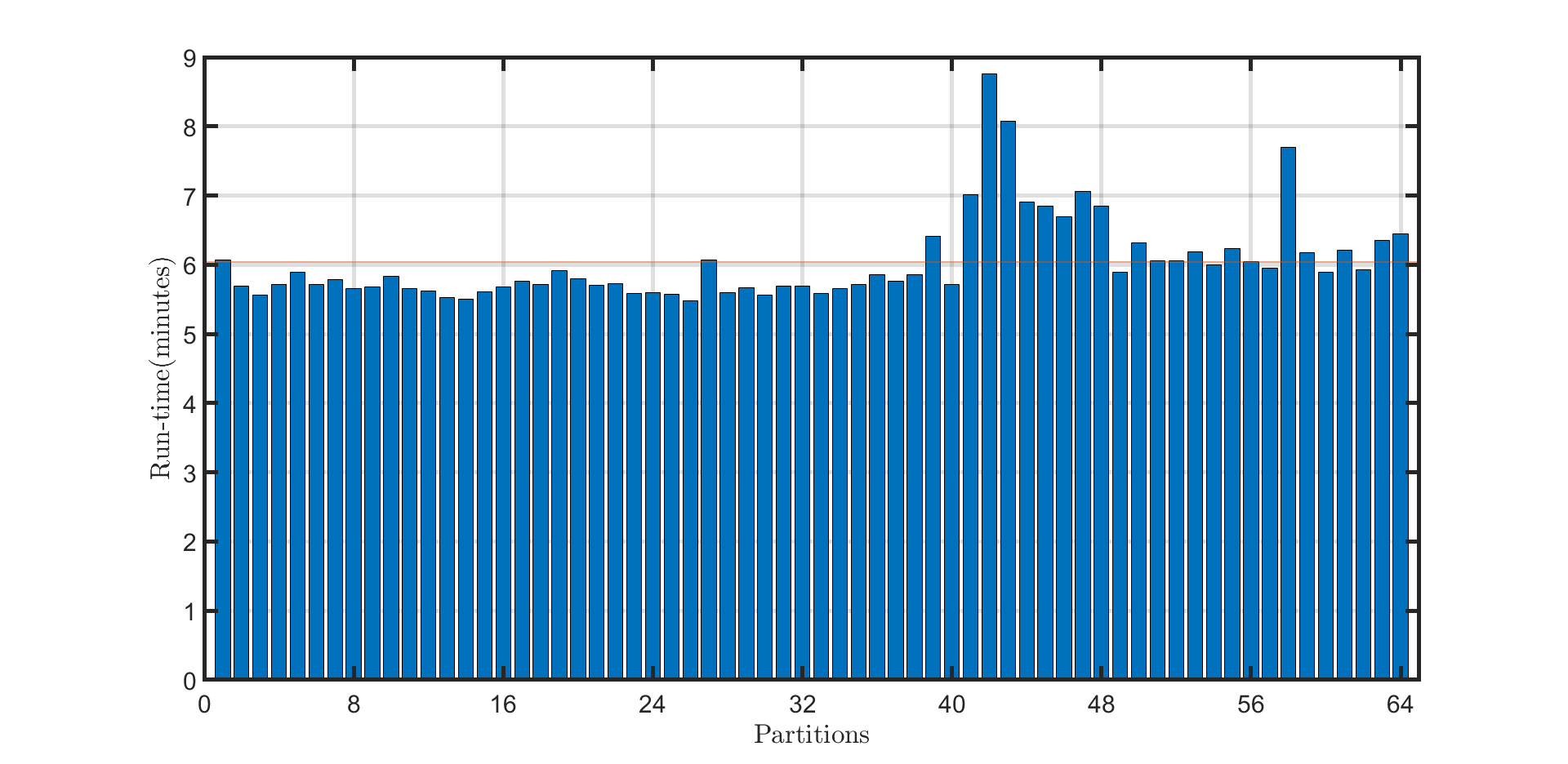}
    \caption{\small{Shows the verification run-time in Linear Time-Varying Plant, for every $64$ partitions of the set of initial states. We apply algorithm \ref{algo:Lipver} on every partition and conclude the verification. The red line shows the average run time on the partitions which is approximately $6$ minutes. On the other hand, the maximum run-time is $8$ minutes and $40$ seconds.}} 
    \label{fig:LTV_runtime}
 \end{figure}

\end{document}